\PassOptionsToPackage{unicode}{hyperref}
\PassOptionsToPackage{hyphens}{url}
\PassOptionsToPackage{dvipsnames,svgnames,x11names}{xcolor}
\documentclass[
]{article}

\usepackage{amsmath,amssymb}
\usepackage{iftex}
\ifPDFTeX
  \usepackage[T1]{fontenc}
  \usepackage[utf8]{inputenc}
  \usepackage{textcomp} 
\else 
  \usepackage{unicode-math}
  \defaultfontfeatures{Scale=MatchLowercase}
  \defaultfontfeatures[\rmfamily]{Ligatures=TeX,Scale=1}
\fi
\usepackage{lmodern}
\ifPDFTeX\else  
\fi
\IfFileExists{upquote.sty}{\usepackage{upquote}}{}
\IfFileExists{microtype.sty}{
  \usepackage[]{microtype}
  \UseMicrotypeSet[protrusion]{basicmath} 
}{}
\makeatletter
\@ifundefined{KOMAClassName}{
  \IfFileExists{parskip.sty}{%
    \usepackage{parskip}
  }{
    \setlength{\parindent}{0pt}
    \setlength{\parskip}{6pt plus 2pt minus 1pt}}
}{
  \KOMAoptions{parskip=half}}
\makeatother
\usepackage{xcolor}
\ifx\paragraph\undefined\else
  \let\oldparagraph\paragraph
  \renewcommand{\paragraph}[1]{\oldparagraph{#1}\mbox{}}
\fi
\ifx\subparagraph\undefined\else
  \let\oldsubparagraph\subparagraph
  \renewcommand{\subparagraph}[1]{\oldsubparagraph{#1}\mbox{}}
\fi

\usepackage{longtable,booktabs,array}
\usepackage{calc} 
\usepackage{etoolbox}
\makeatletter
\patchcmd\longtable{\par}{\if@noskipsec\mbox{}\fi\par}{}{}
\makeatother
\IfFileExists{footnotehyper.sty}{\usepackage{footnotehyper}}{\usepackage{footnote}}
\makesavenoteenv{longtable}
\usepackage{graphicx}
\makeatletter
\def\maxwidth{\ifdim\Gin@nat@width>\linewidth\linewidth\else\Gin@nat@width\fi}
\def\maxheight{\ifdim\Gin@nat@height>\textheight\textheight\else\Gin@nat@height\fi}
\makeatother
\setkeys{Gin}{width=\maxwidth,height=\maxheight,keepaspectratio}
\makeatletter
\def\fps@figure{htbp}
\makeatother
\newlength{\cslhangindent}
\newlength{\csllabelwidth}
\newlength{\cslentryspacingunit} 
\newenvironment{CSLReferences}[2] 
 {
  \setlength{\parindent}{0pt}
  \ifodd #1
  \let\oldpar\par
  \def\par{\hangindent=\cslhangindent\oldpar}
  \fi
  \setlength{\parskip}{#2\cslentryspacingunit}
 }%
 {}
\usepackage{calc}

\usepackage{booktabs}
\usepackage{longtable}
\usepackage{array}
\usepackage{multirow}
\usepackage{wrapfig}
\usepackage{float}
\usepackage{colortbl}
\usepackage{pdflscape}
\usepackage{tabu}
\usepackage{threeparttable}
\usepackage{threeparttablex}
\usepackage[normalem]{ulem}
\usepackage{makecell}
\usepackage{xcolor}
\usepackage{arxiv}
\usepackage{orcidlink}
\usepackage{amsmath}
\usepackage[T1]{fontenc}
\makeatletter
\@ifpackageloaded{caption}{}{\usepackage{caption}}
\AtBeginDocument{%
\ifdefined\contentsname
  \renewcommand*\contentsname{Table of contents}
\else
  \newcommand\contentsname{Table of contents}
\fi
\ifdefined\listfigurename
  \renewcommand*\listfigurename{List of Figures}
\else
  \newcommand\listfigurename{List of Figures}
\fi
\ifdefined\listtablename
  \renewcommand*\listtablename{List of Tables}
\else
  \newcommand\listtablename{List of Tables}
\fi
\ifdefined\figurename
  \renewcommand*\figurename{Figure}
\else
  \newcommand\figurename{Figure}
\fi
\ifdefined\tablename
  \renewcommand*\tablename{Table}
\else
  \newcommand\tablename{Table}
\fi
}
\@ifpackageloaded{float}{}{\usepackage{float}}
\floatstyle{ruled}
\@ifundefined{c@chapter}{\newfloat{codelisting}{h}{lop}}{\newfloat{codelisting}{h}{lop}[chapter]}
\floatname{codelisting}{Listing}

\makeatother
\makeatletter
\@ifpackageloaded{caption}{}{\usepackage{caption}}
\@ifpackageloaded{subcaption}{}{\usepackage{subcaption}}
\makeatother
\makeatletter
\@ifpackageloaded{tcolorbox}{}{\usepackage[skins,breakable]{tcolorbox}}
\makeatother
\makeatletter
\@ifundefined{shadecolor}{\definecolor{shadecolor}{rgb}{.97, .97, .97}}
\makeatother
\ifLuaTeX
  \usepackage{selnolig}  
\fi
\IfFileExists{bookmark.sty}{\usepackage{bookmark}}{\usepackage{hyperref}}
\IfFileExists{xurl.sty}{\usepackage{xurl}}{} 
\hypersetup{
  pdftitle={New allometric models for the USA create a step-change in forest carbon estimation, modeling, and mapping},
  pdfauthor={Lucas K Johnson; Michael J Mahoney; Grant M Domke; Colin M Beier},
  pdfkeywords={aboveground biomass, national forest
inventory, Landsat, design-based estimation, model-based estimation},
  colorlinks=true,
  linkcolor={blue},
  filecolor={Maroon},
  citecolor={Blue},
  urlcolor={Blue},
  pdfcreator={LaTeX via pandoc}}

\renewcommand{\today}{2024-05-07}
\newcommand{\runninghead}{A Preprint }
\renewcommand{\runninghead}{A step-change in forest carbon mapping }
\title{New allometric models for the USA create a step-change in forest
carbon estimation, modeling, and mapping}
\def\asep{\\\\\\ } 
\author{\textbf{Lucas K
Johnson}~\orcidlink{0000-0002-7953-0260}\\Division of Environmental
Science\\State University of New York College of Environmental Science
and Forestry\\Syracuse,
NY,\ 13210\\\href{mailto:ljohns11@esf.edu}{ljohns11@esf.edu}\asep\textbf{Michael
J Mahoney}~\orcidlink{0000-0003-2402-304X}\\Division of Environmental
Science\\State University of New York College of Environmental Science
and Forestry\\Syracuse,
NY,\ 13210\\\href{mailto:mjmahone@esf.edu}{mjmahone@esf.edu}\asep\textbf{Grant
M Domke}~\orcidlink{0000-0003-0485-0355}\\Northern Research
Station\\USDA Forest Service\\St.~Paul,
MN,\ 55108\\\href{mailto:grant.m.domke@usda.gov}{grant.m.domke@usda.gov}\asep\textbf{Colin
M Beier}~\orcidlink{0000-0003-2692-7296}\\Department of Sustainable
Resources Management\\State University of New York College of
Environmental Science and Forestry\\Syracuse,
NY,\ 13210\\\href{mailto:cbeier@esf.edu}{cbeier@esf.edu}}
\date{2024-05-07}
\begin{document}
\maketitle
\begin{abstract}
The United States national forest inventory (NFI) serves as the
foundation for forest aboveground biomass (AGB) and carbon accounting
across the nation. These data enable design-based estimates of forest
carbon stocks and stock-changes at state and regional levels, but also
serve as inputs to model-based approaches for characterizing forest
carbon stocks and stock-changes at finer spatial and temporal
resolutions. Although NFI tree and plot-level data are often treated as
truth in these models, they are in fact estimates based on regional
species-group models and parameters known collectively as the Component
Ratio Method (CRM). In late 2023 the Forest Inventory and Analysis (FIA)
program introduced a new National Scale Volume and Biomass Estimators
(NSVB) system to replace CRM nationwide and offer more precise and
accurate representations of forest AGB and carbon. Given the prevalence
of model-based AGB studies relying on FIA, there is concern about the
transferability of methods from CRM to NSVB models, as well as the
comparability of existing CRM AGB products (e.g.~maps) to new and
forthcoming NSVB AGB products. To begin addressing these concerns we
compared previously published CRM AGB maps and estimates to new maps and
estimates produced using identical methods with NSVB AGB reference data.
Our results suggest that models relying on passive satellite imagery
(e.g.~Landsat) provide acceptable estimates of point-in-time NSVB AGB
and carbon stocks, but fail to accurately quantify growth in mature
closed-canopy forests. We highlight that existing estimates, models, and
maps based on FIA reference data are no longer compatible with NSVB, and
recommend new methods as well as updated models and maps for
accommodating this step-change. Our collective ability to adopt NSVB in
our modeling and mapping workflows will help us provide the most
accurate spatial forest carbon data possible in order to better inform
local management and decision making.
\end{abstract}
{\bfseries \emph Keywords}
\def\sep{\textbullet\ }
aboveground biomass \sep national forest
inventory \sep Landsat \sep design-based estimation \sep 
model-based estimation

\ifdefined\Shaded\renewenvironment{Shaded}{\begin{tcolorbox}[interior hidden, boxrule=0pt, sharp corners, borderline west={3pt}{0pt}{shadecolor}, frame hidden, breakable, enhanced]}{\end{tcolorbox}}\fi

\hypertarget{introduction}{%
\section{Introduction}\label{introduction}}

Measuring and monitoring forest biomass is an active area of research,
with mapped predictions of forest aboveground biomass (AGB) in
particular being used to inform carbon accounting efforts alongside
other conservation and stewardship activities. In the US context, such
maps often rely upon AGB estimates from the US Forest Inventory and
Analysis program (FIA) which provides tree-level AGB estimates for
millions of trees across thousands of plots throughout the US (Gray et
al. 2012). These tree-level estimates can then be aggregated to plots
and used to estimate AGB across the landscape (Bechtold and Patterson
2005). However, these design-based estimates are limited to a relatively
coarse spatial resolution due to the density of the FIA sample
(McRoberts 2011). A plethora of studies have demonstrated using
model-based estimation to produce AGB estimates at finer resolutions,
fitting models to predict FIA's AGB estimates using remote sensing data
products (for instance, Kennedy, Ohmann, et al. 2018; Johnson et al.
2022, 2023; Huang et al. 2019; Ayrey et al. 2021; Zheng, Heath, and
Ducey 2007). These studies commonly treat the AGB estimates provided by
FIA as ground truth and adopt them with only minimal adjustments or
corrections (if any).

However, FIA's AGB estimates are not actual ground truth measurements,
but are instead calculated by applying allometric models to quantities
more easily measured in the field. Historically these estimates have
been calculated using the component ratio method (CRM, Woodall, Heath,
et al. 2011; Heath et al. 2009), a combination of regional
species-group-specific allometric models that estimate biomass as a
function of tree diameter and height. Although AGB estimates from CRM
are useful, they do have several limitations. In particular, CRM models
for non-merchantable portions of trees were not constructed using actual
tree component measurements and the administrative boundaries used to
delineate regional models did not correspond to underlying ecological
gradients.

In light of these limitations and the rapidly growing interest in
measuring and monitoring AGB and forest carbon stocks, FIA introduced a
new National Scale Volume and Biomass Estimators (NSVB) system to
replace CRM nationwide in late 2023 (Westfall et al. 2023). NSVB
estimates of AGB address the primary concerns with CRM and provide a
more unified and accurate set of estimates for AGB across the nation.
NSVB also represents a step-change in forest AGB estimation, with NSVB
estimating 14.6\% more forest AGB nationwide than CRM estimates using
the same field measurements. This difference varies across geographic
regions and forest compositions with differences in estimated
state-level AGB ranging from +37.1\% for Indiana to -1.5\% for
Washington (Westfall et al. 2023; Virginia Tech 2023a, 2023c). A major
driver of these differences appears to be improvements to the estimation
of top and limb biomass, and as a result species that are poor
self-pruners or otherwise grow denser tops and branches often (though
not always) saw the largest increases in estimated AGB under NSVB.

The new AGB and carbon estimates introduced with NSVB do not reflect
on-the-ground changes in forest structure or function, but simply
represent accounting changes resulting from swapping one system of
allometric models for another. Although they are purely structural,
these accounting changes aim to improve our understanding of forest AGB
and carbon stocks and fluxes across the landscape by increasing the
precision and accuracy of tree-level volume, biomass, and carbon
estimates. Therefore it is necessary to update existing
carbon-monitoring systems that rely on FIA estimates of AGB or carbon to
incorporate the benefits of NSVB in resulting map products and
estimates. However, given the importance of top and limb biomass in NSVB
estimates, it is not clear if common model-based approaches relying on
remotely sensed data (which may have difficulty penetrating the canopy
to capture the density of tops and branches in a forest) will be
effective at predicting NSVB estimates. This in turn has obvious
downstream implications for how forest carbon accounting will be
impacted by the shift from CRM to NSVB estimates of forest AGB and
carbon stocks, and suggests the urgent need to revisit existing modeling
pipelines holistically rather than exchanging individual components
(e.g. reference data sets) in isolation. Further, there exists concern
about the comparability of existing maps produced using CRM AGB to newer
products produced using NSVB AGB. It is not yet known whether AGB maps
produced using NSVB will be simple ``rescalings'' of their CRM
equivalents, or if these new maps will exhibit new patterns across focal
landscapes. These questions and concerns are particularly pressing given
the proliferation of AGB models based on FIA data and both the growing
demand and technical capacity for remote monitoring of forest carbon
stocks.

To begin answering these questions, we compared previously published AGB
maps fit using CRM AGB estimates to new maps produced using identical
methods and NSVB AGB estimates. Specifically, we 1) replicated the
methodology of Johnson et al. (2023), a well-established empirical
modeling approach relying on field inventory plots and Landsat imagery
(Lu 2006; Hall et al. 2006; Powell et al. 2010; Pflugmacher, Cohen, and
E. Kennedy 2012; Deo et al. 2016; Kennedy, Ohmann, et al. 2018), to
produce maps of AGB based on NSVB. 2) We compared these new NSVB-based
AGB maps against CRM-based maps from Johnson et al. (2023) for 2005 and
2019. 3) We then quantified the agreement between these maps, the
differences in their AGB stock-change estimates for 2005-2019, the
spatial patterns of these differences, and the ability for machine
learning models fit using Landsat imagery to estimate CRM and NSVB AGB
overall.

\hypertarget{methods}{%
\section{Methods}\label{methods}}

\hypertarget{sec-studyarea}{%
\subsection{Study area}\label{sec-studyarea}}

New York State covers 141,297 \(\operatorname{km}^2\) of the
northeastern United States. Daily temperatures across the state ranged
from -17 \(^\circ\)C to 28 \(^\circ\)C in 2019, while monthly
precipitation for the same period ranged from 5.0 cm to 16.8 cm (NOAA
National Centers for Environmental Information 2022). Elevation ranges
from -2 m to 1,584 m above sea level (U.S. Geological Survey 2019). The
majority of the state occupies the northern hardwoods-hemlock forest
region, though there are important inclusions of beech-maple-basswood
and Appalachian oak communities in the western and southern reaches of
the state, respectively (Dyer 2006). Like much of the US Northeast, NYS
was extensively deforested during the 18th and 19th centuries, with
subsequent reforestation and conservation resulting in a landscape
dominated by forest stands that are now over 100 years old (Whitney
1994; Lorimer 2001; Mahoney et al. 2022). NYS created the Forest
Preserve in 1885, establishing the foundation for what became the
Adirondack and Catskill Parks decades later. Any state-owned or acquired
lands within these parks has since been designated as `forever wild' and
has largely been preserved, with no timber harvesting.

The US Forest Service estimates that NSVB-induced changes to AGB
estimates for New York State will be moderate, with an estimated 5.7\%
increase in AGB statewide (Virginia Tech 2023b). This increase is not
distributed evenly across species, however; FIA estimates that red
spruce (\emph{Picea rubens}) AGB will increase by approximately 38.7\%,
while sugar maple (\emph{Acer saccharum}) AGB will increase by just
under 5\%, and yellow birch (\emph{Betula alleghaniensis}) AGB will
decrease by roughly 5\% (Virginia Tech 2023b). These species are not
uniformly distributed throughout the state: red spruce is most common in
high-elevation forests throughout the Adirondack Park, sugar maple is
the largest contributor to statewide total standing biomass (Virginia
Tech 2023b) with highest abundance in montane regions with more mature
forests, often in combination with yellow birch (Wilson, Lister, and
Riemann 2012; Riemann et al. 2014). As a result, it is not yet known how
local to regional estimates of forest AGB and carbon sequestration will
be impacted by the shift to NSVB.

\hypertarget{modeling-nsvb-agb}{%
\subsection{Modeling NSVB AGB}\label{modeling-nsvb-agb}}

We repeated the methods used in the ``direct'' modeling procedure from
Johnson et al. (2023) to develop AGB maps for New York State using NSVB
AGB estimates and Landsat imagery from 2005 and 2019 as summarized
below.

\hypertarget{sec-field-inv}{%
\subsubsection{Field inventory data}\label{sec-field-inv}}

Each FIA plot contains four 7.32 m radius (24 ft) circular subplots, one
at the center of the plot and three centered 36.6 m (120 ft) away from
the plot center at azimuths of 120\(^{\circ}\), 240\(^{\circ}\), and
360\(^{\circ}\) (Bechtold and Patterson 2005). We obtained true plot
centroids for FIA plots within New York State under an agreement with
the FIA program office. Plots are assigned to one of five panels, each
assumed to have complete spatial coverage across the state, and
remeasured on a 5-7 year basis. Trees on land considered to be forested
within each subplot above 12.7 cm (5 in) DBH are measured and assigned
AGB estimates. We aggregated tree-level AGB estimates to the FIA
plot-level, and area-normalized these estimates to convert them into
units of \(\operatorname{Mg\ ha}^{-1}\). FIA considers land to be
forested if an area at least 0.4 ha in size and 120m in width is not
developed for nonforest land uses and is at least 10\% stocked with
trees. Under this definition, it is likely that some woody AGB exists on
some of these nonforest lands not measured by FIA. However, in the
absence of field measurements for these areas, we treated these
nonforest lands as having 0 AGB.

We filtered the complete set of FIA plots within New York State measured
between 2002 and 2019 to only those inventories where all four subplots
were measured. When plots in this filtered set were inventoried multiple
times, we randomly selected a single inventory year to avoid
pseudoreplication. This initial selection criteria resulted in a pool of
5,144 plots. We then divided this set of plots into model development
and map assessment datasets using FIA's panel designation, with one of
the five panels randomly selected and all plots with this designation
assigned to the map assessment dataset, and the remaining plots assigned
to the model development dataset. In this way we partitioned 20\% of the
available data for an independent map assessment, yielding a probability
sample with complete spatial coverage which allowed us to use
design-unbiased estimators for map agreement metrics (Riemann et al.
2010; Stehman and Foody 2019).

We then filtered the model development set to only include completely
forested (and therefore completely measured) plots. We then added to
this set a group of completely nonforested plots with maximum
LiDAR-derived canopy heights \(\leq\) 1m, which we assumed to correspond
to having 0 AGB (Johnson et al. 2022), resulting in a total of 2,049
model development plots. The map assessment set was further filtered to
remove plots external to our mapped area based on our landcover mask
(Johnson et al. 2023), as these plots were outside our population of
interest, resulting in 545 assessment plots.

\hypertarget{auxiliary-data}{%
\subsubsection{Auxiliary data}\label{auxiliary-data}}

Following Mahoney et al. (2022), we derived a set of 16 predictors from
annual growing-season medoid composites of Landsat collection 1 imagery
(USGS 2018) using coefficients from Roy et al. (2016). These predictors
were then processed using Landtrendr in Google Earth Engine using
normalized burn ratio temporally segmented vertices to produce a
smoothed continuous 30-year time series of pixel-level metrics (Kennedy,
Yang, and Cohen 2010; Kennedy, Yang, et al. 2018). Metrics derived
included normalized burn ratio (Kauth and Thomas 1976), tasseled-cap
wetness, brightness, and greenness (Cocke, Fulé, and Crouse 2005),
normalized difference vegetation index (Kriegler et al. 1969), simple
ratio (Jordan 1969), and modified simple ratio (Chen 1996), as well as
the one-year change at each pixel for each of these indices. Index
calculation used the spectral module for Google Earth Engine (Montero et
al. 2022). We additionally calculated the years since the most recent
disturbance from the normalized burn ratio segmentation, with a temporal
range from 1985-2019, and the associated magnitude of change in
normalized burn ratio.

To this set of auxiliary data we added annual primary and secondary land
cover classifications from the US Geological Survey's Land Change
Monitoring, Assessment and Projection (LCMAP) version 1.2 (Zhu and
Woodcock 2014; Brown et al. 2020) and a set of steady-state variables
including 30 year normals for precipitation and temperature (PRISM
Climate Group 2022), elevation, aspect, slope, and topographic wetness
index derived from a 30 m digital elevation model (Beven and Kirkby
1979; U.S. Geological Survey 2019; Mahoney, Beier, and Ackerman 2022),
estimates from a 2005 global canopy height model (Simard et al. 2011;
Hudak et al. 2020), distance to nearest waterbodies identified by the US
Census Bureau (US Census Bureau 2013; Walker 2023), National Wetland
Inventory classifications from the US Fish and Wildlife Service, and the
Environmental Protection Agency's level 4 ecozones (Omernik and Griffith
2014; CEC 1997). Ecozones were aggregated to their corresponding level 3
ecozone if they did not cover \(\geq\) 2\% of the state. If the
aggregated level 3 ecozones also did not cover \(\geq\) 2\% of the
state, they were assigned an ecozone of ``other''.

All auxiliary data was projected to match Landsat 30 m pixel geometries,
using the terra pacakge (Hijmans 2023) for the R programming language (R
Core Team 2023). We used the exactextractr library to associate the
gridded auxiliary data with FIA plots (described in
Section~\ref{sec-field-inv}), calculating the mean of pixel values
intersecting with subplot locations weighted by the fraction of each
pixel overlapping with the plot area (Baston 2022).

\hypertarget{model-fitting}{%
\subsubsection{Model fitting}\label{model-fitting}}

We further partitioned the model development data set (described in
Section~\ref{sec-field-inv}) into a training set (80\% of observations)
and a testing set (20\% of observations). We fit a random forest as
implemented in the ranger R package (Breiman 2001; Wright and Ziegler
2017), a gradient boosting machine as implemented in the lightgbm R
package (Friedman 2002; Ke et al. 2017; Shi et al. 2022), and a support
vector machine as implemented in the kernlab R package (Cortes and
Vapnik 1995; Karatzoglou et al. 2004) to this training set.
Hyperparameters were selected for each method through an iterative grid
search using five-fold cross-validation. More information on this tuning
procedure is available in Section 2.6 (Model development) of Johnson et
al. (2023).

We then used five-fold cross-validation to generate predictions from
each of these models, using their selected hyperparameters, for each
observation in the training data set. We then fit a linear regression
estimating plot AGB as a function of these predictions, ensembling the
three base models in order to produce a single final prediction (Wolpert
1992; Dormann et al. 2018). We selected this ensemble model as our final
model; all further discussions of model accuracy and predictions refer
to this ensemble. Following Johnson et al. (2023), and in order to avoid
redundant comparisons between models of NSVB and CRM AGB, we focused
only on the map accuracy assessment and have provided model testing set
metrics in Supplementary Materials A.

\hypertarget{mapping-and-postprocessing}{%
\subsection{Mapping and
postprocessing}\label{mapping-and-postprocessing}}

We used the ensemble model to make predictions for all 30 m pixels
across the state for both 2019, the most recent map year in Johnson et
al. (2023), and 2005, the earliest year possible to produce a statewide
design-based estimate from the public FIA database. Following Johnson et
al. (2023), we masked our predictions with annual LCMAP version 1.2
primary landcover classification product to remove developed, cropland,
water, and barren pixels. We then compared these maps to CRM-based maps
representing 2019 and 2005 from Johnson et al. (2023).

\hypertarget{sec-map-acc}{%
\subsection{Map accuracy assessment}\label{sec-map-acc}}

As this study represents one of the first (if not the first) model-based
estimation studies based on NSVB AGB, we perform a comprehensive map
accuracy assessment of our NSVB AGB model following methods described in
Riemann et al. (2010). Using the ensemble model, we predicted AGB at
each 30m pixel coincident with FIA plots, using predictors derived from
Landsat imagery from the same year as the FIA inventory. We then used
the exactextractr library to calculate the mean of these pixel
predictions at each plot, weighted by the fraction of each pixel
overlapping with the plot area (Baston 2022).

Following Riemann et al. (2010) we calculated model assessment metrics
both directly using these weighted averages and FIA plot-level estimates
(plot-to-pixel) and after aggregating predictions and FIA plot-level
estimates to multiple regular grids of variably-sized hexagons,
increasing in size such that hexagon centroids ranged from 2 km to 50 km
apart (1 km separation yields the equivalent to a plot-to-pixel
comparison).

Model assessment metrics included root-mean squared error (RMSE;
Equation~\ref{eq-rmse}), mean absolute error (MAE;
Equation~\ref{eq-mae}), mean error (ME; Equation~\ref{eq-me}), and
\(\operatorname{R^2}\) (Equation~\ref{eq-r2}):

\begin{equation}\protect\hypertarget{eq-rmse}{}{
\operatorname{RMSE} = \sqrt{(\frac{1}{n})\sum_{i=1}^{n}(y_{i} - \hat{y_{i}})^{2}}
}\label{eq-rmse}\end{equation}

\begin{equation}\protect\hypertarget{eq-mae}{}{
\operatorname{MAE} = (\frac{1}{n})\sum_{i=1}^{n}\left | y_{i} -\hat{y_{i}} \right |
}\label{eq-mae}\end{equation}

\begin{equation}\protect\hypertarget{eq-me}{}{
\operatorname{ME} = (\frac{1}{n})\sum_{i=1}^{n}y_{i} -\hat{y_{i}}
}\label{eq-me}\end{equation}

\begin{equation}\protect\hypertarget{eq-r2}{}{
\operatorname{R^2} = 1 - \frac{\sum_{i=1}^{n}\left(y_{i}-\hat{y}_{i}\right)^2}{\sum_{i=1}^{n}\left(y_i - \bar{y}\right)^2}
}\label{eq-r2}\end{equation}

Where \(y_{i}\) is FIA estimated AGB, \(\hat{y}_{i}\) is model-predicted
AGB, and \(\bar{y}\) is mean FIA estimated AGB.

Interpreting RMSE and MAE can be difficult, as these metrics are in
units of \(\operatorname{Mg\ ha}^{-1}\) which can make interpret the
relative magnitude of errors challenging without also knowing the mean
and distribution of pixel AGB values. To aid interpretation, we present
normalized variants of these metrics rescaled by the mean AGB within the
training data set:

\begin{equation}\protect\hypertarget{eq-rmsecv}{}{
\operatorname{RMSE\ \%} = 100 \cdot \frac{\operatorname{RMSE}}{\bar{y}}
}\label{eq-rmsecv}\end{equation}

\begin{equation}\protect\hypertarget{eq-maecv}{}{
\operatorname{MAE\ \%} = 100 \cdot \frac{\operatorname{MAE}}{\bar{y}}
}\label{eq-maecv}\end{equation}

These metrics are well-suited to describing a single model, or for
comparing multiple models fit to the same data set. However, they are
not well suited for comparing models fit to outcomes pulled from
different distributions. In order to compare the performance of models
fit to NSVB AGB estimates to the CRM AGB model from Johnson et al.
(2023), we instead used the dimensionless \(d_r\) statistic from
Willmott, Robeson, and Matsuura (2011):

\begin{equation}\protect\hypertarget{eq-willmott-dr}{}{
d_r = \left\{\begin{array}{l}1 - \dfrac{\sum_{i=1}^{n}{\left|\hat{y}_{i}-y_{i}\right|}}{c\sum_{i=1}^{n}{\left|y_{i}-\bar{y}\right|}}, \text{ when}\\\\ \sum_{i=1}^{n}{\left|\hat{y}_{i}-y_{i}\right|} \leq c\sum_{i=1}^{n}{\left|y_{i}-\bar{y}\right|}\\\\\dfrac{c\sum_{i=1}^{n}{\left|y_{i}-\bar{y}\right|}}{\sum_{i=1}^{n}{\left|\hat{y}_{i}-y_{i}\right|}} - 1 \text{ otherwise}\end{array}\right\}
}\label{eq-willmott-dr}\end{equation}

The \(d_r\) metric is bounded \([-1, 1]\), with a value of 1 indicating
perfect agreement between \(y\) and \(\hat{y}\). It is an asymmetrical
metric which assumes that \(y\) are ``true'' values, which is
appropriate for comparing model predictions to FIA estimates. As such,
we calculated \(d_r\) for both the model fit using NSVB AGB and the
model fit using CRM AGB, using the corresponding set of FIA AGB
estimates, in order to compare how well these models represent FIA AGB
estimates.

Each of these metrics was calculated at a plot-to-pixel scale and after
aggregation to each grid of hexagons. Metric calculations used the
waywiser R package (Mahoney 2023).

Finally, using the same set of map assessment plots, we assessed how
well the distribution of predictions from CRM and NSVB AGB models
approximated the distributions of their corresponding FIA estimates
using empirical cumulative distribution function plots and
Kolmogorov-Smirnov (KS) tests (Massey Jr 1951).

\hypertarget{map-comparisons}{%
\subsection{Map comparisons}\label{map-comparisons}}

We directly compared a 2019 NSVB AGB map produced in this study to a
2019 CRM AGB map produced by the direct modeling approach in Johnson et
al. (2023) through raster differencing. We differenced AGB maps and
percent rank maps to identify spatial patterns of both absolute changes
between CRM and NSVB predictions as well as shifts relative to the
statewide distributions. We also assessed the agreement between these
2019 AGB maps across multiple scales of aggregation with variably-sized
hexagons (centroids 1 km to 50 km apart; Section~\ref{sec-map-acc};
Riemann et al. (2010)) using the agreement coefficient (AC;
Equation~\ref{eq-ac}) from Ji and Gallo (2006):

\begin{equation}\protect\hypertarget{eq-ac}{}{
\operatorname{AC} = 1 - \frac{\sum_{i=1}^{n}{\left(\hat{y}_{i}-y_{i}\right)^{2}}}{\sum_{i=1}^{n}{\left(\left|\bar{\hat{y}}-\bar{y}\right|+\left|\hat{y}_{i}-\bar{\hat{y}}\right|\right)\left(\left|\bar{\hat{y}}-\bar{y}\right|+\left|y_{i}-\bar{y}\right|\right)}}
}\label{eq-ac}\end{equation}

Here \(y\) represents predictions from the CRM model, while \(\hat{y}\)
represents predictions from the NSVB model. AC is a symmetrical metric
which allows for errors in both \(y\) and \(\hat{y}\), making it useful
to compare model predictions to one another. AC is bounded
\([-\infty, 1]\), with a value of 1 representing perfect agreement.

A benefit of AC is that it can be broken down into systematic and
unsystematic components by first estimating a geometric mean functional
relationship (GMFR) model (Equation~\ref{eq-gmfr}) to predict \(y\) as a
function of \(\hat{y}\) (Draper and Yang 1997):

\begin{equation}\protect\hypertarget{eq-gmfr}{}{
\begin{aligned}
y^\prime{} &= a + b\hat{y} \\
b &= \pm \left(\frac{\sum_{i=1}^{n}{\left(\hat{y}_{i}-\bar{\hat{y}}\right)^{2}}}{\sum_{i=1}^{n}{\left({y}_{i}-\bar{{y}}\right)^{2}}}\right)^{\frac{1}{2}} \\
a &= \bar{\hat{y}} - b\bar{y}
\end{aligned}
}\label{eq-gmfr}\end{equation}

Where the sign of \(b\) is the same as the correlation coefficient
between \(y\) and \(\hat{y}\), and \(y^{\prime}\) represents
GMFR-predicted values of \(y\).

We then use the GMFR to predict \(\hat{y}\) as a function of \(y\),
which we term \(\hat{y}^{\prime}\):

\begin{equation}\protect\hypertarget{eq-gmfr-flip}{}{
\hat{y}^\prime{} = -\frac{a}{b} + \frac{1}{by}
}\label{eq-gmfr-flip}\end{equation}

We then use \(\hat{y}^{\prime}\) to decompose the sum of squared
differences in the numerator of AC (Equation~\ref{eq-ac}), into its
unsystematic (\(\operatorname{AC}_u\), Equation~\ref{eq-acu}) and
systematic (\(\operatorname{AC}_s\), Equation~\ref{eq-acs}) components:

\begin{equation}\protect\hypertarget{eq-acu}{}{
\operatorname{AC}_u = 1 - \frac{\sum_{i=1}^{n}{\left[\left(\left|\hat{y}_{i}-\hat{y}_i^\prime{}\right|\right)\left(\left|{y}_{i}-{y}_i^\prime{}\right|\right)\right]}}{\sum_{i=1}^{n}{\left(\left|\bar{\hat{y}}-\bar{y}\right|+\left|\hat{y}_{i}-\bar{\hat{y}}\right|\right)\left(\left|\bar{\hat{y}}-\bar{y}\right|+\left|y_{i}-\bar{y}\right|\right)}}
}\label{eq-acu}\end{equation}

\begin{equation}\protect\hypertarget{eq-acs}{}{
\operatorname{AC}_s = 1 - \frac{\left(\sum_{i=1}^{n}{\left(\hat{y}_{i}-y_{i}\right)^{2}}\right) - \left(\sum_{i=1}^{n}{\left[\left(\left|\hat{y}_{i}-\hat{y}_i^\prime{}\right|\right)\left(\left|{y}_{i}-{y}_i^\prime{}\right|\right)\right]}\right)}{\sum_{i=1}^{n}{\left(\left|\bar{\hat{y}}-\bar{y}\right|+\left|\hat{y}_{i}-\bar{\hat{y}}\right|\right)\left(\left|\bar{\hat{y}}-\bar{y}\right|+\left|y_{i}-\bar{y}\right|\right)}}
}\label{eq-acs}\end{equation}

In this way, we were able to determine the degree to which differences
in maps produced using CRM AGB are systematically different from maps
produced using NSVB AGB, and therefore the degree to which they could be
linearly rescaled to match NSVB-based estimates. Further details of this
derivation are provided in Ji and Gallo (2006).

We tested rescaling model-based CRM AGB to model-based NSVB AGB with a
simple linear regression as follows:

\begin{equation}\protect\hypertarget{eq-rescale}{}{
\operatorname{NSVB_{mod}} = \beta_{0} + \beta_{1} \cdot CRM_{mod} + \beta_{2} \cdot Elevation
}\label{eq-rescale}\end{equation}

Where \(\operatorname{NSVB_{mod}}\), \(\operatorname{CRM_{mod}}\), and
Elevation were pixel-level values sampled from a raster stack containing
maps of NSVB and CRM AGB (2019; \(\operatorname{Mg\ ha^{-1}}\)) and
elevation (m) as layers. We randomly divided the 1,000,000 pixel sample
into training (80\%; n = 800,000) and testing (20\%; n = 200,000) sets
for model fitting and assessment respectively.

\hypertarget{sec-stock}{%
\subsection{Statewide stock estimates}\label{sec-stock}}

We additionally compared estimates of statewide AGB stocks calculated
across both sets of estimation methods (design-based vs model-based
approaches) and allometric models (CRM vs NSVB). We estimated total NYS
AGB and aboveground carbon (AGC) stocks for 2005 and 2019 using
design-based estimates following methods described in Pugh et al.
(2018), using FIA inventory data with both NSVB and CRM-based AGB
estimates. These estimates included stems 12.7 cm (5 inches) and up and
correspond to all land in NYS. We then calculated model-based AGB stocks
for both the CRM and NSVB-based models by multiplying the average
pixel-level AGB density (\(\operatorname{Mg\ ha}^{-1}\)) by the total
area of NYS. Following CRM protocols described in Woodall, Heath, et al.
(2011), AGC stocks were assumed to be 50\% of AGB for all model-based
CRM estimates. Under the new NSVB protocols described in Westfall et al.
(2023), species-specific carbon fractions are used to convert AGB to
AGC. We did not have access to a fine-resolution species map for NYS in
our map years, but instead computed statewide carbon fractions for each
respective map-year (2005 and 2019) as an average of species-specific
carbon fractions, weighted by each species' relative contribution to
FIA's statewide AGB estimates. We used these fractions to convert
model-based estimates of statewide total NSVB AGB to statewide total
NSVB AGC.

\hypertarget{sec-stockchngmeth}{%
\subsection{Comparing stock-changes with NSVB and
CRM}\label{sec-stockchngmeth}}

We computed statewide design-based and Landsat-based stock-change
estimates by subtracting the 2005 estimates for each method from their
corresponding 2019 estimates. Additionally, we compared pixel-level
stock-changes by first computing stock change maps as 2019 AGB minus
2005 AGB for NSVB and CRM and then differencing these stock change maps
(\(\Delta\) NSVB - \(\Delta\) CRM). We used this stock-change difference
map to identify landscape patterns of differences in statewide
stock-change estimates.

\hypertarget{results}{%
\section{Results}\label{results}}

\hypertarget{nsvb-map-accuracy}{%
\subsection{NSVB Map Accuracy}\label{nsvb-map-accuracy}}

\begin{figure}

{\centering \includegraphics{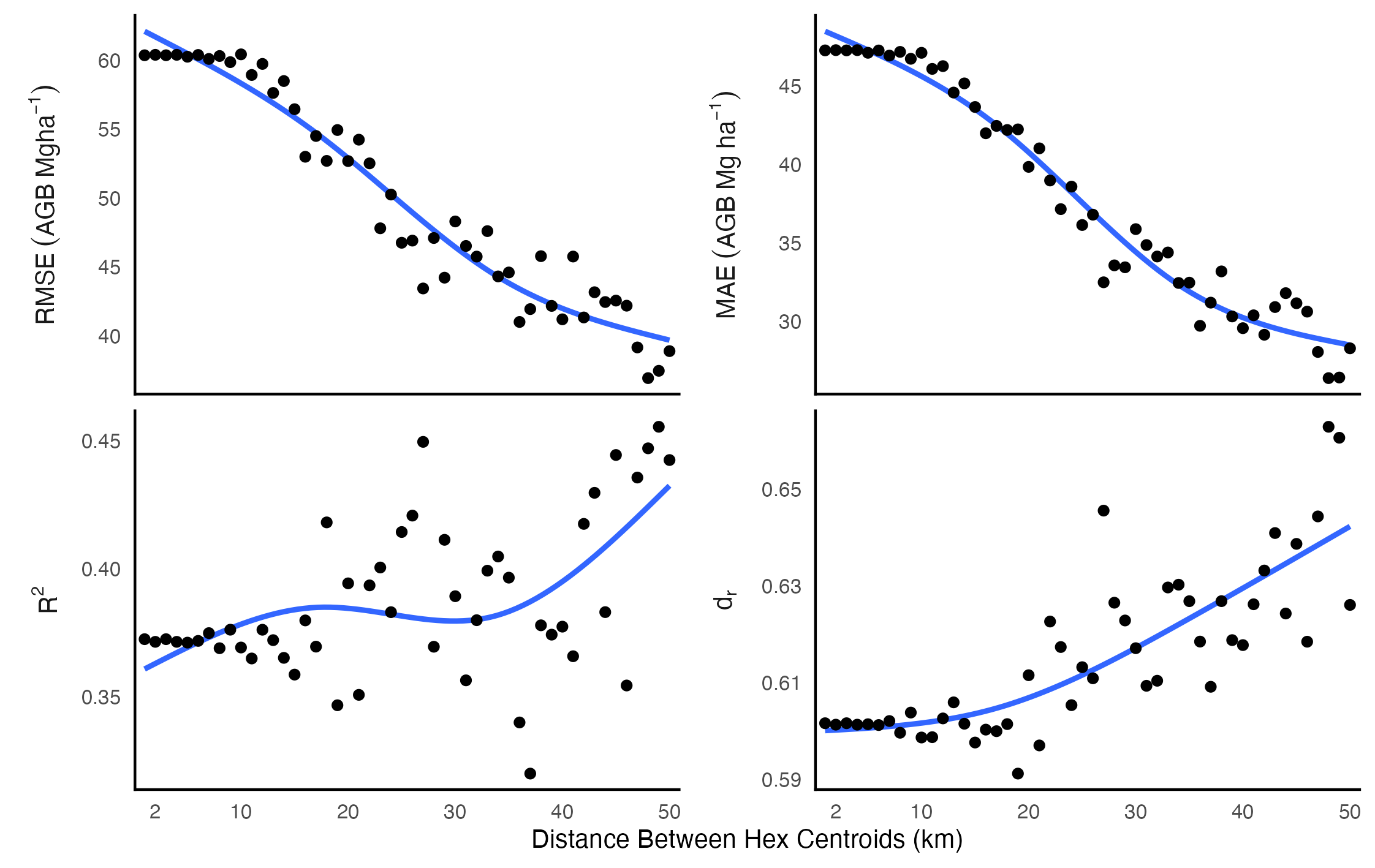}

}

\caption{\label{fig-nsvbacc}NSVB AGB map accuracy results (vs FIA)
across multiple scales represented by distances between hexagon
centroids. Blue lines are fit to data using cubic splines with three
knots.}

\end{figure}

Mapped AGB predictions from the NSVB model became increasingly accurate
and exhibited higher agreement with FIA estimates as aggregation unit
size increased, with \% MAE decreasing from 36.01 to 20.93\%, \% RMSE
decreasing from 45.97 to 28.81\%, \(\operatorname{R^{2}}\) increasing
from 0.37 to 0.44, and \(\operatorname{d_{r}}\) increasing from 0.60 to
0.63 (Table~\ref{tbl-nsvbacc}; Figure~\ref{fig-nsvbacc}). RMSE and MAE
followed similar patterns. ME estimates were small across all scales of
assessment and did not exhibit a monotonic improvement with aggregation
(Table~\ref{tbl-nsvbacc}).

\hypertarget{tbl-nsvbacc}{}
\begin{table}
\caption{\label{tbl-nsvbacc}NSVB AGB map accuracy results (vs FIA) for select scales. RMSE, MAE, ME
in \(\operatorname{Mg\ ha}^{-1}\). Scale = distance between hexagon
centroids in km; PPH = plots per hexagon; n = number of comparison units
(plots or hexagons). All metrics as defined in Section 2.3. }\tabularnewline

\centering
\begin{tabular}[t]{lrrrrrrrrr}
\toprule
\multicolumn{1}{c}{Scale} & \multicolumn{1}{c}{n} & \multicolumn{1}{c}{PPH} & \multicolumn{1}{c}{MAE} & \multicolumn{1}{c}{\% MAE} & \multicolumn{1}{c}{RMSE} & \multicolumn{1}{c}{\% RMSE} & \multicolumn{1}{c}{ME} & \multicolumn{1}{c}{$\operatorname{R^2}$} & \multicolumn{1}{c}{$\operatorname{d_r}$}\\
\midrule
Plot:Pixel & 545 &  & 47.25 & 36.01 & 60.33 & 45.97 & -3.86 & 0.37 & 0.60\\
\addlinespace
20 km & 293 & 1.86 & 39.82 & 30.03 & 52.66 & 39.72 & -2.32 & 0.39 & 0.61\\
\addlinespace
30 km & 171 & 3.19 & 35.85 & 26.99 & 48.28 & 36.34 & -0.72 & 0.39 & 0.62\\
\addlinespace
50 km & 74 & 7.38 & 28.24 & 20.93 & 38.87 & 28.81 & 3.39 & 0.44 & 0.63\\
\bottomrule
\end{tabular}
\end{table}

\hypertarget{nsvb-vs-crm-map}{%
\subsection{NSVB vs CRM Map}\label{nsvb-vs-crm-map}}

\begin{figure}

{\centering \includegraphics{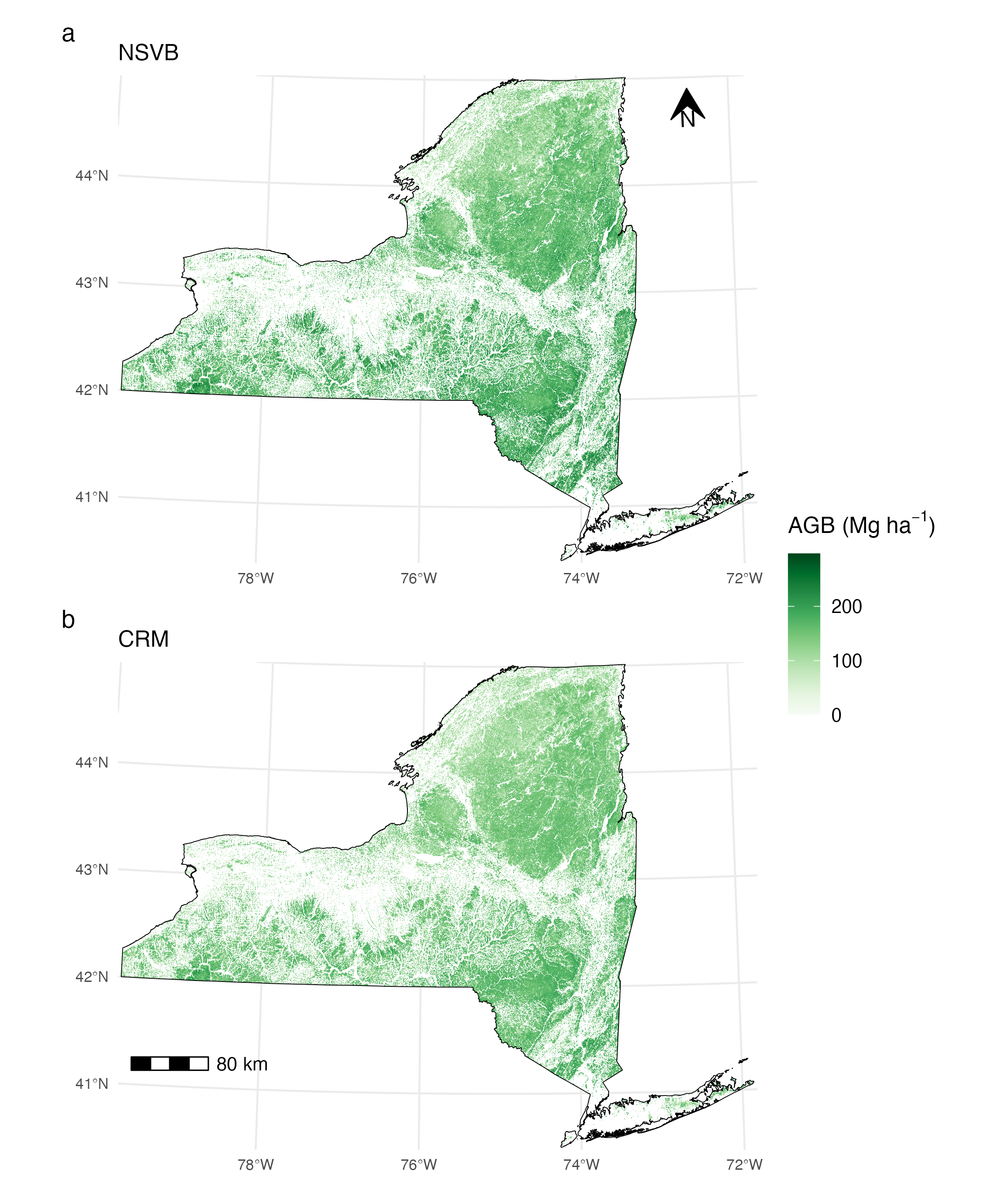}

}

\caption{\label{fig-agb2019}Landsat modeled AGB for 2019. a) Predictions
made with models trained on NSVB-based FIA reference data. b)
Predictions made with models trained on CRM-based FIA reference data.}

\end{figure}

\begin{figure}

{\centering \includegraphics{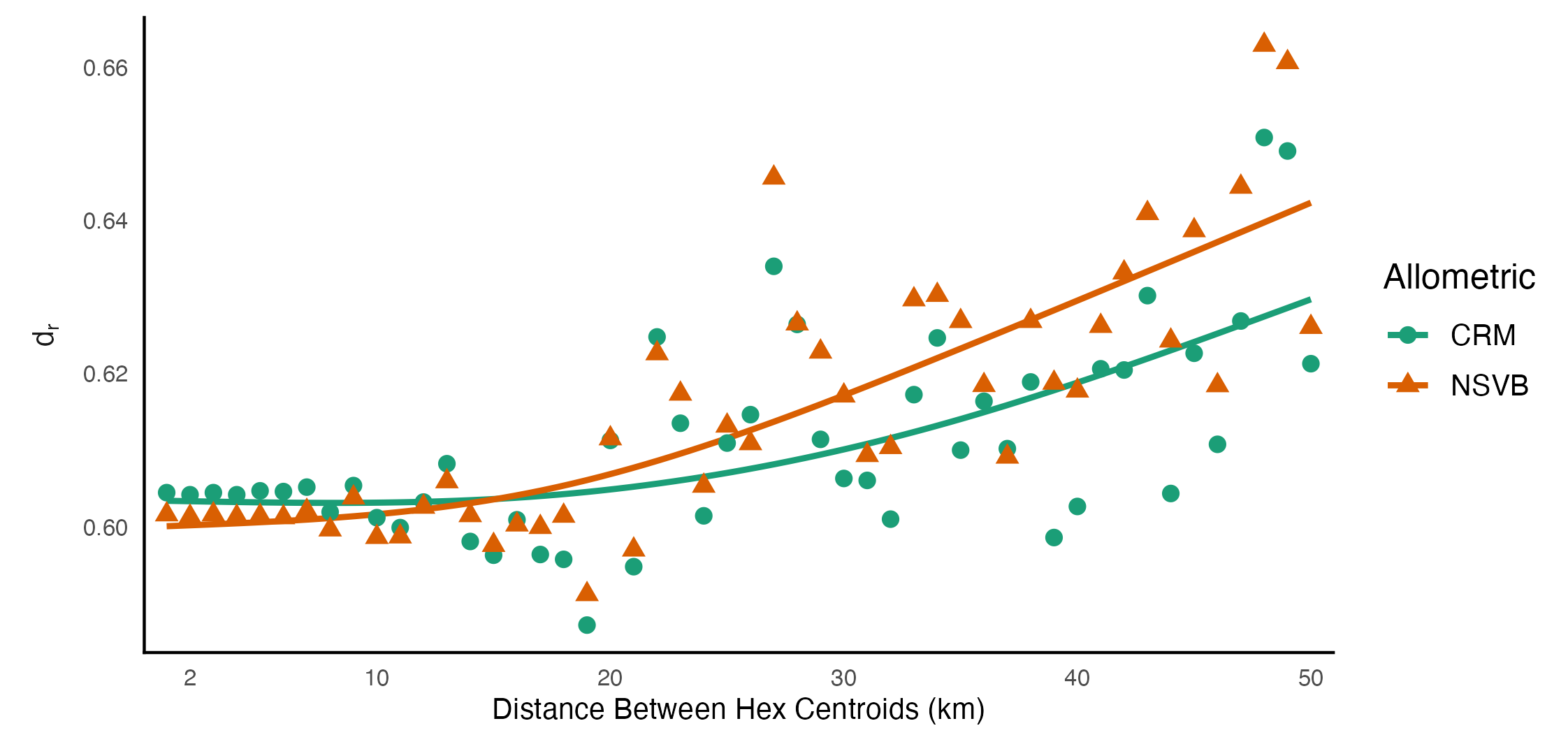}

}

\caption{\label{fig-dr}Willmott's \(\operatorname{d_{r}}\) (measure of
agreement, 1 is perfect; Section 2.4) across scales represented by
distances between hexagon centroids. Landsat-based NSVB AGB vs FIA-based
NSVB AGB (orange) and Landsat-based CRM AGB vs FIA-based CRM AGB
(green). Trend lines fit to data using cubic splines with three knots.}

\end{figure}

\begin{figure}

{\centering \includegraphics{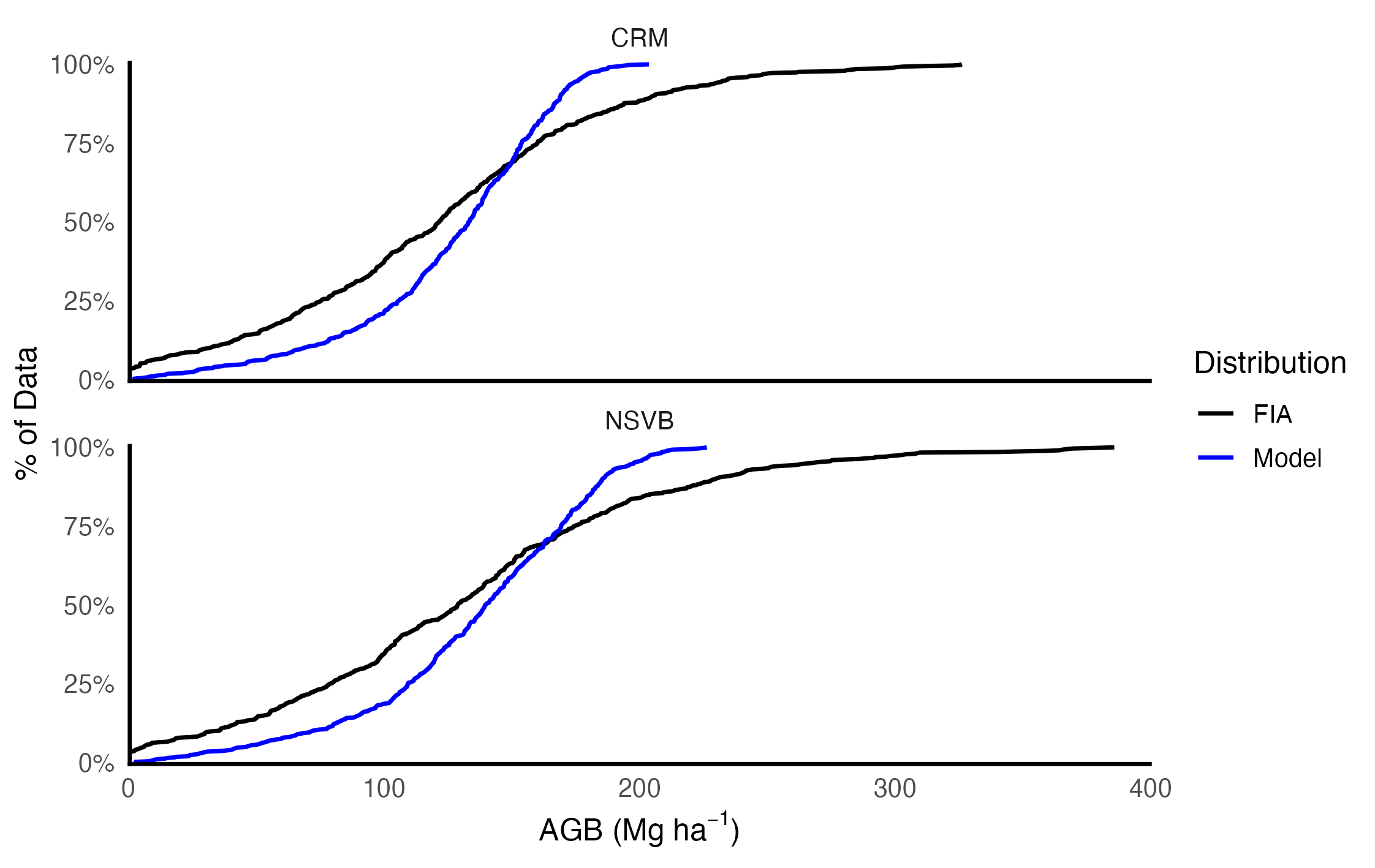}

}

\caption{\label{fig-ecdf}Empirical cumulative distribution function
comparisons. Landsat-modeled predictions of AGB vs FIA plot-level
estimates of AGB for NSVB and CRM respectively.}

\end{figure}

Both NSVB and CRM maps (Figure~\ref{fig-agb2019}) yielded similar
agreement with corresponding FIA estimates across all scales of
aggregation; the dimensionless \(\operatorname{d_{r}}\) statistics
(ranging from -1 to 1) differed by at most 0.02 between the approaches
(Figure~\ref{fig-dr}). CRM and NSVB AGB models approximated the
distributions of their corresponding FIA estimates similarly
(Figure~\ref{fig-ecdf}), but NSVB models appeared be more impacted by
saturation, a common issue when modeling forest structure with passive
satellite imagery (Lu 2005; Duncanson, Niemann, and Wulder 2010; Johnson
et al. 2023), where models underpredict the largest reference
observations. Kolmogorov-Smirnov statistics confirmed that CRM AGB
models were only marginally better at approximating the corresponding
FIA distribution (CRM 0.17; NSVB 0.18).

\begin{figure}

{\centering \includegraphics{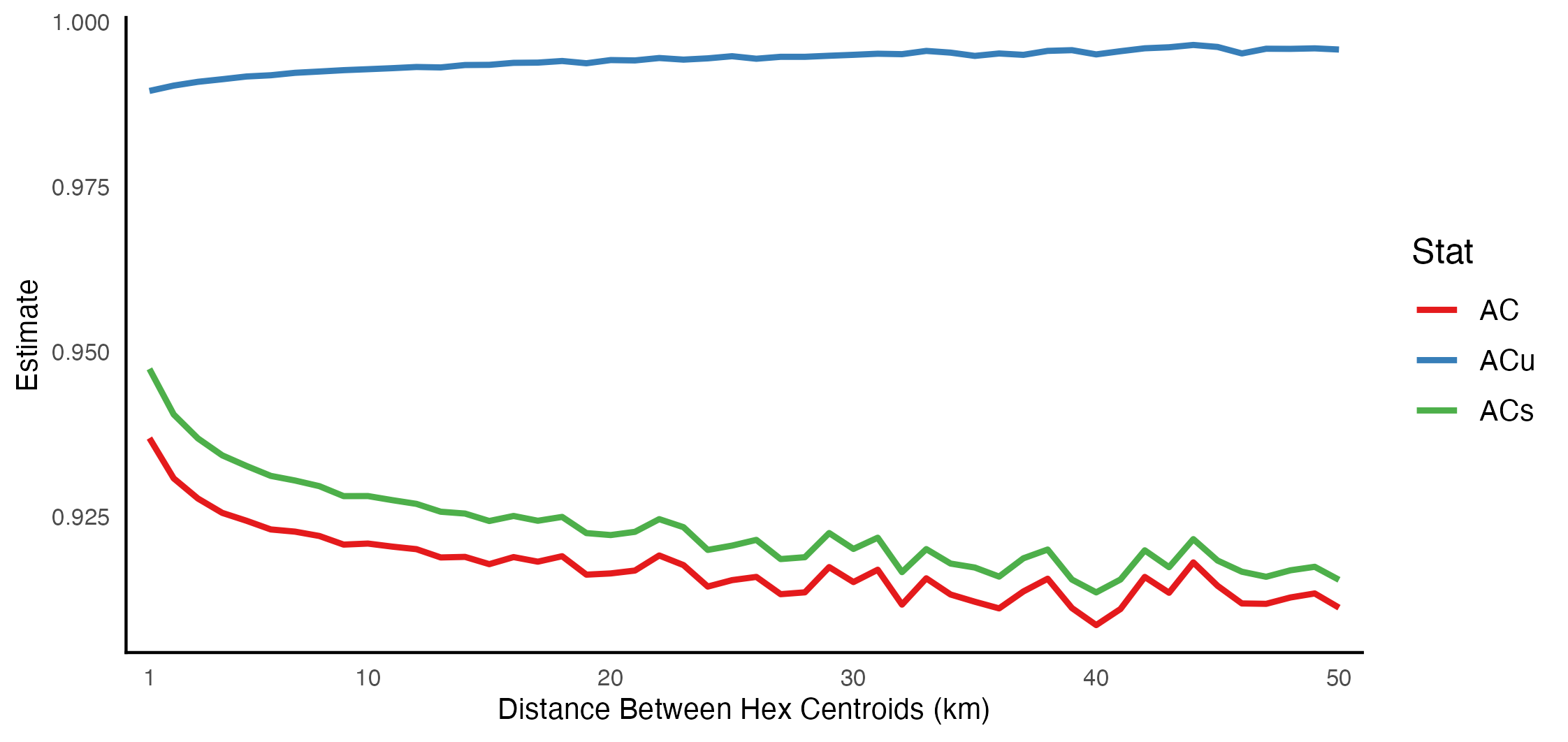}

}

\caption{\label{fig-ac}Agreement between NSVB AGB maps and CRM AGB maps
across scales represented by distances between hexagon centroids. AC
(agreement coefficient) and its systematic (ACs) and unsystematic (ACu)
components as defined in Section 2.5.}

\end{figure}

\begin{figure}

{\centering \includegraphics{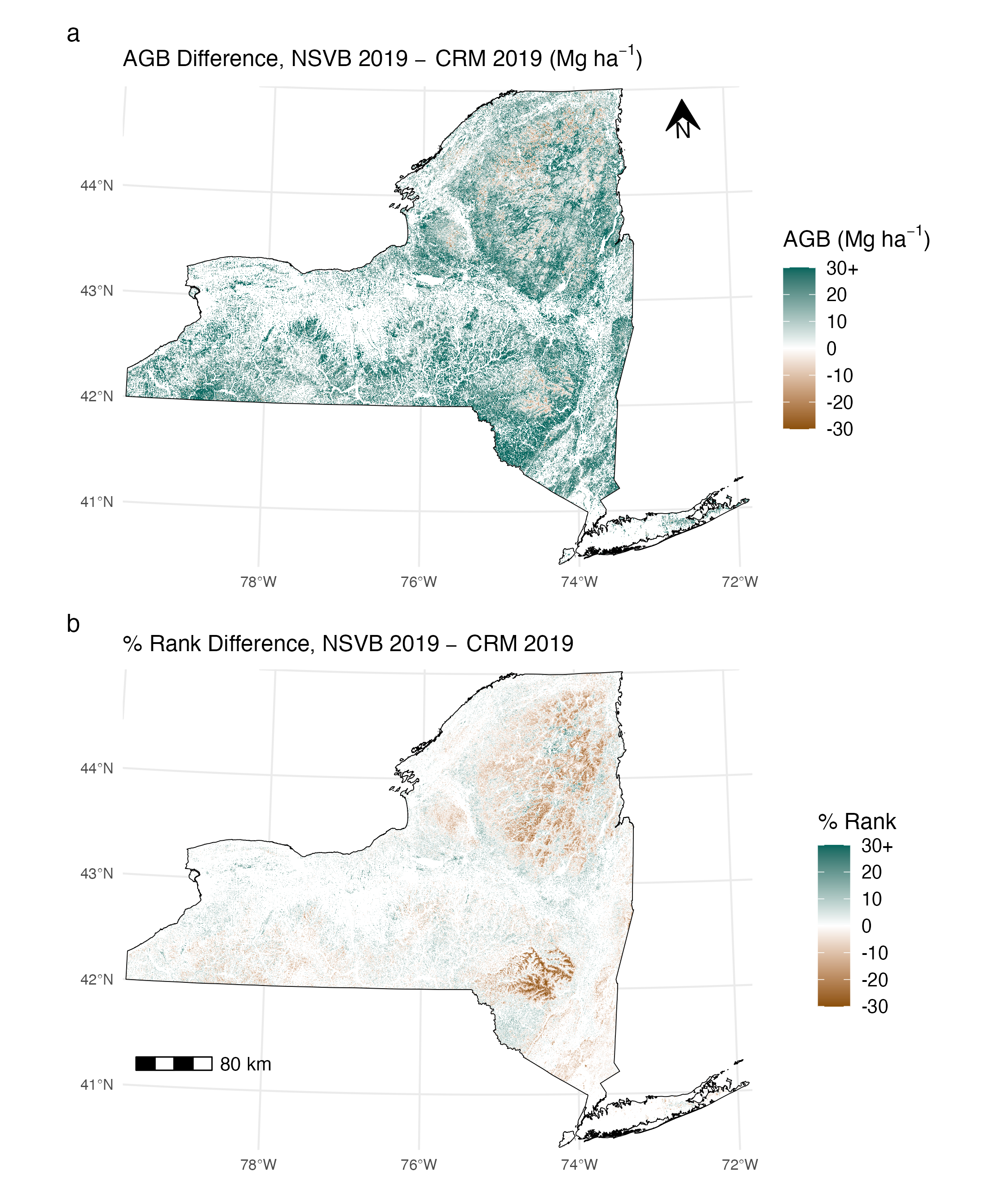}

}

\caption{\label{fig-diffmap}Difference maps for 2019 computed as
Landsat-modeled AGB with NSVB-based FIA reference data minus
Landsat-modeled AGB with CRM-based FIA reference data. a) AGB
differences in units of \(\operatorname{Mg\ ha}^{-1}\) b) Percent rank
differences. Mapped values capped at +/- 30.}

\end{figure}

NSVB and CRM AGB maps agreed well with one another across all scales of
aggregation (\(\operatorname{AC > 0.9}\)), but agreement worsened with
increasing aggregation (Figure~\ref{fig-ac}). The majority of the
disagreement was due to systematic differences between the maps (ACs).
The AGB difference map showed that models of NSVB AGB produced larger
estimates relative to models of CRM AGB across most of NYS, but produced
smaller estimates concentrated in what appear to be the montane forests
of the Adirondack and Catskill regions (Figure~\ref{fig-diffmap} a). In
contrast to the AGB difference map, the percent rank difference map
showed a more uniform pattern of downshifts and upshifts in percent rank
across NYS, highlighting `gaining' and `losing' areas in terms of AGB
density (more or less AGB relative to the rest of the state) with a
shift to NSVB. High-elevation areas still contained the largest relative
downshifts in percent rank from CRM to NSVB (Figure~\ref{fig-diffmap} b)
indicating that models estimated these areas to be much less AGB-dense
relative to the rest of NYS following a change from CRM to NSVB.

The linear regression (Equation~\ref{eq-rescale}) used to rescale CRM
AGB to NSVB AGB was accurate (RMSE 14.05, MAE 10.91, ME -0.01) and
represented 93\% (\(\operatorname{R^{2}}\)) of the variability in the
model-based NSVB predictions from the test set. The regression
coefficients succinctly described the observed spatial patterns of
differences between model-based CRM and NSVB AGB
(Figure~\ref{fig-diffmap}), indicating that NSVB AGB would increase by
\textasciitilde1.14 \(\operatorname{Mg\ ha^{-1}}\) with every unit
increase in CRM AGB and elevation held constant, but that NSVB AGB would
decrease by \textasciitilde2 \(\operatorname{Mg\ ha^{-1}}\) for every
100 m of elevation gained and CRM AGB held constant
(Table~\ref{tbl-rescale}; Supplementary Materials B).

\hypertarget{tbl-rescale}{}
\begin{table}
\caption{\label{tbl-rescale}Results of a regression model representing model-based NSVB AGB
(\(\operatorname{Mg\ ha}^{-1}\)) as a function of model-based CRM AGB
(\(\operatorname{Mg\ ha}^{-1}\)) and elevation (m). }\tabularnewline

\centering
\begin{tabular}[t]{lrrrr}
\toprule
\multicolumn{1}{c}{ } & \multicolumn{1}{c}{Coefficient} & \multicolumn{1}{c}{Standard error} & \multicolumn{1}{c}{t} & \multicolumn{1}{c}{p}\\
\midrule
Intercept & 9.555 & 0.050 & 189.855 & < 0.001\\
\addlinespace
CRM AGB & 1.135 & < 0.001 & 3,322.265 & < 0.001\\
\addlinespace
Elevation & -0.023 & < 0.001 & -275.981 & < 0.001\\
\bottomrule
\end{tabular}
\end{table}

\hypertarget{stock-and-stock-change-comparison}{%
\subsection{Stock and stock-change
comparison}\label{stock-and-stock-change-comparison}}

Estimates of total NSVB AGB and AGC for NYS were substantially larger
than corresponding CRM estimates for both the design-based and
model-based methods (Table~\ref{tbl-stocks}). However, while
design-based stock-change estimates were greater when using NSVB
allometrics, model-based stock-change estimates using NSVB were smaller
than their CRM counterparts. Additionally, there was greater divergence
between design-based and model-based estimates of both statewide AGB and
AGC when using NSVB instead of CRM (Table~\ref{tbl-stocks}).

\begin{figure}

{\centering \includegraphics{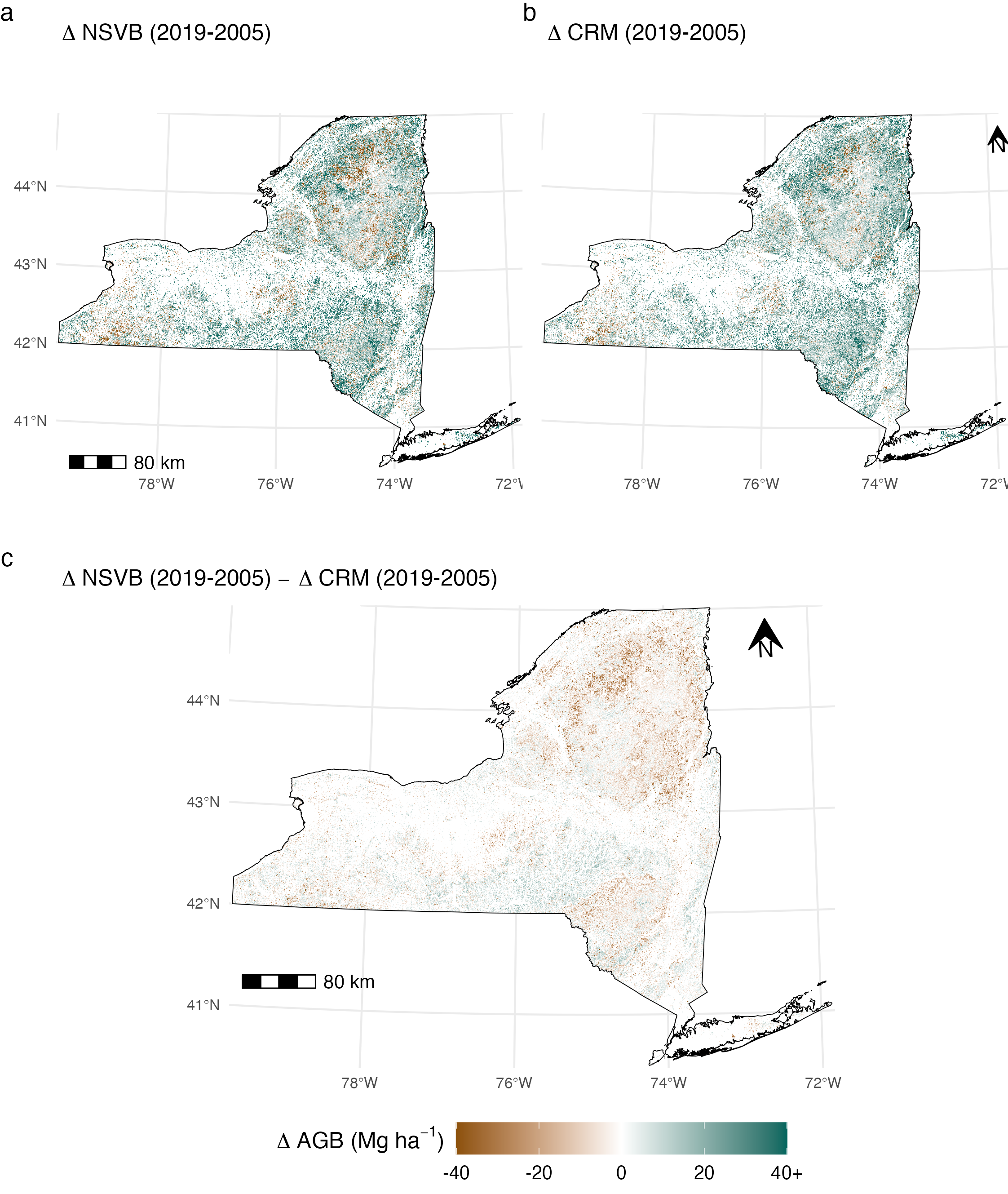}

}

\caption{\label{fig-stockchange}Stock-change map comparison. a) 15-year
NSVB stock-change map (\(\Delta\) NSVB; 2019 AGB - 2005 AGB). b) 15-year
CRM stock-change map (\(\Delta\) CRM; 2019 AGB - 2005 AGB). c)
Stock-change difference map computed as \(\Delta\) NSVB minus \(\Delta\)
CRM. Mapped values capped at +/- 40 \(\operatorname{Mg\ ha^{-1}}\).}

\end{figure}

The model-based stock-change maps for NSVB and CRM yielded nearly
identical spatial distributions of AGB losses and gains, however the
magnitudes of losses and gains differed (Figure~\ref{fig-stockchange} a,
b). AGB losses from 2005 to 2019 appeared to be greater in magnitude in
the NSVB map than in the corresponding CRM map, particularly in the
Adirondack region in the northeastern portion of the state
(Figure~\ref{fig-stockchange} c). NSVB AGB gains over the time period
were only marginally larger than corresponding CRM AGB gains throughout
the south-central region, and along the southeastern border of the
state.

\hypertarget{tbl-stocks}{}
\begin{table}
\caption{\label{tbl-stocks}Statewide stocks (2019, 2005) and stock-changes (\(\Delta\); 2019 minus
2005) for aboveground biomass (AGB) and aboveground carbon (AGC) by
estimation method (FIA vs Landsat) and allometric (CRM vs NSVB). Values
in millions of metric tons. }\tabularnewline

\centering
\begin{tabular}[t]{llrrrrrr}
\toprule
\multicolumn{2}{c}{ } & \multicolumn{3}{c}{CRM} & \multicolumn{3}{c}{NSVB} \\
\cmidrule(l{3pt}r{3pt}){3-5} \cmidrule(l{3pt}r{3pt}){6-8}
\multicolumn{1}{c}{} & \multicolumn{1}{c}{} & \multicolumn{1}{c}{FIA} & \multicolumn{1}{c}{Landsat} & \multicolumn{1}{c}{FIA - Landsat} & \multicolumn{1}{c}{FIA} & \multicolumn{1}{c}{Landsat} & \multicolumn{1}{c}{FIA - Landsat}\\
\midrule
 & 2005 & 910.29 & 943.01 & -32.72 & 967.42 & 1106.10 & -138.68\\

 & 2019 & 1038.87 & 1063.90 & -25.03 & 1129.59 & 1213.03 & -83.44\\

\multirow{-3}{*}{\raggedright\arraybackslash AGB} & $\Delta$ & 128.58 & 120.90 & 7.68 & 162.17 & 106.93 & 55.24\\
\cmidrule{1-8}
 & 2005 & 455.14 & 471.50 & -16.36 & 467.89 & 534.87 & -66.98\\

 & 2019 & 519.44 & 531.95 & -12.51 & 546.13 & 586.69 & -40.56\\

\multirow{-3}{*}{\raggedright\arraybackslash AGC} & $\Delta$ & 64.29 & 60.45 & 3.84 & 78.24 & 51.81 & 26.43\\
\bottomrule
\end{tabular}
\end{table}

\hypertarget{discussion}{%
\section{Discussion}\label{discussion}}

In this study we investigated the transferability of well-established
methods for estimating, modeling, and mapping aboveground biomass (AGB)
based on Landsat imagery between two USDA Forest Inventory and Analysis
(FIA) reference data sets constructed with different systems of
allometric models: the legacy Component Ratio Method (CRM, Woodall,
Domke, et al. 2011) and the new National Scale Volume and Biomass
Estimators (NSVB, Westfall et al. 2023). To this end, we compared
previously published CRM AGB maps and derived estimates (Johnson et al.
2023) to new maps and derived estimates produced using identical methods
but with NSVB AGB reference data. We found that NSVB and CRM AGB maps
agreed well with one another, though systematic differences driven by
NSVB allometric equations were apparent in our maps; NSVB AGB
predictions were generally larger than corresponding CRM AGB predictions
and concentrations of smaller NSVB AGB existed in high-elevation regions
of the state. Maps of NSVB and CRM AGB agreed similarly with
corresponding FIA reference data sets, however models of NSVB AGB were
more impacted by saturation and thus were limited in their ability to
characterize AGB growth beyond a certain point. These findings
demonstrate that NSVB represents a step-change for model-based forest
carbon estimation and monitoring in the United States (US) and suggest
the need to update entire modeling pipelines, especially for
applications that require the representation of forest growth over time.
In particular incorporating NSVB-based reference data in carbon
monitoring systems may require updated models, careful comparisons with
CRM-based estimates, and new modeling approaches including additional
remotely sensed datasets such as LiDAR.

\hypertarget{comparing-nsvb-and-crm-agb-maps}{%
\subsection{Comparing NSVB and CRM AGB
maps}\label{comparing-nsvb-and-crm-agb-maps}}

We interpreted that the differences between our NSVB and CRM AGB maps
(Figure~\ref{fig-diffmap}) were driven by changes in the underlying
system of allometric models, specifically how CRM and NSVB estimates
differed for individual species and size classes. NSVB was designed (in
part) to yield more precise estimates of top and limb biomass and as a
result, the majority of the differences between the estimates from the
two systems can be attributed to these non-merchantable components of
trees (Westfall et al. 2023; Virginia Tech 2023b). Within NYS, an FIA
estimated 57.5\% increase in statewide top and limb biomass contributed
to an overall 5.7\% increase in total AGB (Virginia Tech 2023b). This
increase was generally reflected in our map differences where NSVB AGB
model predictions were larger than corresponding CRM AGB predictions
across most of the state. NSVB AGB increases relative to CRM AGB may be
most pronounced for large-diameter individuals and for species that are
poor self-pruners or otherwise grow densely-packed branches (Westfall et
al. 2023; Virginia Tech 2023b). Conversely, trees in high-elevation
areas where we observed smaller predictions of NSVB AGB relative to CRM
AGB (Table~\ref{tbl-rescale}; Figure~\ref{fig-diffmap}) often have
smaller stems due to shallow soils and more frequent windthrow and ice
storms. Thus, it stands to reason that high-elevation forests generally
would not see the same increase in top and branch biomass with the
transition to NSVB as more size- and species-diverse forests throughout
the state. These same montane regions contain large proportions of sugar
maple and yellow birch (Wilson, Lister, and Riemann 2012; Riemann et al.
2014) which were among the species that did not show large AGB increases
associated with the transition to NSVB (Virginia Tech 2023b). These map
differences exemplify the degree to which updates and improvements to
allometric models may yield structural changes to downstream model-based
estimates along relevant ecological gradients such as species
composition, climate, and topography.

Although there were notable differences between maps
(Figure~\ref{fig-diffmap}), statistical comparisons indicated a high
degree of agreement between NSVB and CRM AGB maps (Figure~\ref{fig-ac}).
Agreement (AC, ACs) decreased as a function of aggregation unit size,
indicating that differences were spatially clustered (particularly in
the aforementioned high-elevation forests), and thus were amplified
through aggregation. Across all scales of aggregation, most of the
disagreement between mapped predictions was attributed to systematic
differences, due to the relatively uniform increase in AGB under NSVB in
most forested regions of the state. As a result, it was possible to
translate existing CRM AGB estimates to their NSVB equivalents through
relatively simple linear relationships (Table~\ref{tbl-rescale};
Equation~\ref{eq-rescale}). However, any such adjustments outside the
scope of this study will require further investigation, will likely be
model- and task-specific, and should be performed with care given the
spatial variability of NSVB-CRM differences.

\hypertarget{difficulty-modeling-nsvb-agb-growth-with-passive-remote-sensing}{%
\subsection{Difficulty modeling NSVB AGB growth with passive remote
sensing}\label{difficulty-modeling-nsvb-agb-growth-with-passive-remote-sensing}}

While the relative increase in FIA-estimated stocks from CRM to NSVB
were mirrored with model-based estimates, the relative decrease in
model-based estimates of stock-changes from CRM to NSVB contradicted
design-based estimates which showed greater stock-changes under NSVB
(Table~\ref{tbl-stocks}). This significantly widened the gap between
design- and model-based stock-change estimates from 7.68 million metric
tons of AGB under CRM, to 55.24 million metric tons of AGB when using
NSVB, calling into question the ability of Landsat-based models to
quantify NSVB stock-changes and forest growth. The introduction of
species-specific carbon fractions with NSVB appeared to have minimal
impact on aboveground carbon estimates relative to the impact of
differences in AGB.

We did find that NSVB AGB models fit using Landsat imagery were still
able to generate acceptable point-in-time AGB stock estimates as
evidenced by the comparable agreement with FIA-reference plots in our
map accuracy assessment (Figure~\ref{fig-dr}). Further, this study does
not cast doubt on the ability of NSVB AGB models to detect discrete
canopy changes driven by deforestation, reforestation, and disturbance
events. However, under NSVB, a greater number of plots had AGB values
beyond the saturation point for model predictions, indicating that
models of NSVB AGB were less capable of accurately estimating AGB in
denser forest stands with larger individual trees than models of CRM AGB
(Figure~\ref{fig-ecdf}). This may be caused by the increased precision
and magnitude of NSVB biomass estimates for the tops and limbs of trees
(Westfall et al. 2023), particularly for larger-diameter trees. These
components are difficult to ``see from space'', particularly in forests
with closed canopies where this biomass is largely hidden from passive
satellite imagery.

Our results therefore suggest that passive remote sensing alone may not
be capable of monitoring NSVB AGB growth, especially in regions similar
to NYS and the much of the Northeastern US where historical land-use
dynamics indicate that the majority of forest stands have either reached
or are approaching maturity (Section~\ref{sec-studyarea}). The inclusion
of active remote sensing systems in NSVB modeling pipelines, including
aerial and spaceborne LiDAR or synthetic aperture radar, may offer the
only solutions to monitoring NSVB AGB growth at scale, as these classes
of sensors have the capacity to penetrate closed canopies and better
represent structural attributes related to AGB, thus reducing saturation
effects (Quegan et al. 2019; Dubayah et al. 2020; Johnson et al. 2022).
Using repeat aerial LiDAR collections to estimate NSVB AGB growth would
be a first step in this direction.

\hypertarget{implications-for-carbon-accounting}{%
\subsection{Implications for carbon
accounting}\label{implications-for-carbon-accounting}}

FIA plot data has been widely used as reference data in model-based
approaches to forest carbon accounting in the US (e.g, Kennedy, Ohmann,
et al. 2018; Johnson et al. 2022, 2023; Huang et al. 2019; Ayrey et al.
2021; Zheng, Heath, and Ducey 2007). Given the reported changes
associated with NSVB (Westfall et al. 2023), it is perhaps unsurprising
that our results show significant differences between model-based NSVB
AGB maps and estimates and their CRM counterparts. However, we have
revealed the potential spatial variability of differences within a
state, uncovered challenges with model-based monitoring of NSVB AGB over
time, and have confirmed the role of species and size-classes in
creating regions of forest carbon gains and losses under a shift to
NSVB. In light of these challenges, the spatial distribution of
differences, and the general lack of consensus on accounting protocols
among carbon markets, there may be incentives to act in bad faith. Of
particular concern is the opportunity to deliberately choose one system
over another on a case-by-case basis in order to maximize payments or
credits for avoided deforestation, aforestation, or improved management
(Gillenwater et al. 2007; Charnley, Diaz, and Gosnell 2010; Kaarakka et
al. 2021). We therefore underscore that this step-change represents an
advance towards a more realistic representation of forest carbon density
across the landscape and thus should be adopted as quickly as possible.
Rescaling existing model-based estimates of CRM AGB to NSVB equivalents
(Equation~\ref{eq-rescale}; Table~\ref{tbl-rescale}) may serve as a
temporary solution to understand how estimates will change or how
decisions may have been different under NSVB. However, it is crucial
that models and maps of CRM AGB are retrained and updated to make the
NSVB transition, especially if we aspire towards map products that have
the most possible practical relevance for local management and carbon
accounting.

Another factor to consider in the NSVB transition is the accurate
conversion of biomass to carbon stocks. NSVB introduced species-specific
carbon fractions, offering improved AGC accuracy over the 0.5 AGB to AGC
ratio assumed with CRM, but at the cost of increased complexity. Under
CRM it was standard practice to map AGB and simply convert these initial
pixel-level predictions to AGC or belowground carbon (BGC) using
fractions that applied to all species (Woodall, Heath, et al. 2011;
Heath et al. 2009). We followed this standard approach in this study for
the sake of comparison to AGB maps produced in Johnson et al. (2023) and
only converted our estimates to AGC for estimates of statewide totals
using a species-weighted carbon fraction (Section~\ref{sec-stock}).
However, under NSVB this AGB-first approach carries the additional
requirement of species maps with matching resolution, extent, and
general timestamp in order to correctly convert each pixel prediction
(Riemann et al. 2014; Wilson, Lister, and Riemann 2012; Wilson et al.
2013). Alternatively, practitioners may opt to model AGC (or BGC)
directly, choosing to incorporate species-specific carbon fractions into
reference data at the plot-level. Stand maps, or regional species-group
maps, may provide a middle road where pixel-level AGB predictions could
be aggregated to stands or regions with approximately known species
compositions, and more targeted species-weighted carbon ratios could be
applied as we did here for statewide estimates of AGC.

\hypertarget{conclusion}{%
\section{Conclusion}\label{conclusion}}

This study represents a first-of-its-kind investigation into using the
new National Scale Volume and Biomass Estimators (NSVB) from the USDA
Forest Inventory and Analysis (FIA) program with existing model-based
approaches to map forest aboveground biomass (AGB) and carbon. Our
results suggest that models relying on passive satellite imagery alone
(e.g.~Landsat) provide acceptable estimates of point-in-time NSVB AGB
and carbon stocks, but fail to appropriately quantify growth in mature
closed-canopy forests. With increasing pressure for forests to sequester
carbon and offset greenhouse gas emissions from other sectors, models
that under-predict or mischaracterize forest growth will not serve to
maximize the role forested landscapes will play in natural climate
solutions and policies. We highlight that existing maps based on FIA
reference data are no longer compatible with NSVB, and recommend
incorporating active remote sensing data, retraining models, rethinking
AGB to carbon conversions, and updating maps to accommodate this
step-change. Doing so is a necessity if we want to meet the growing need
for accurate spatial forest carbon data that can inform management and
decision making across scales.

We can interpret the good practices for greenhouse gas inventories
outlined in the 2019 refinement (Buendia et al. 2019) to the 2006
Intergovernmental Panel on Climate Change (IPCC) guidelines (Eggleston
et al. 2006) as ``a set of procedures intended to ensure that greenhouse
gas inventories are accurate in the sense that they are systematically
neither over- nor underestimates so far as can be judged, and that they
are precise so far as practicable.'' Thus, we have a clear obligation
from the IPCC to update our GHG modeling pipelines with improved methods
and data streams as they become available. However, we have demonstrated
that an improvement to one component of a modeling pipeline in isolation
(e.g.~improved precision and accuracy in reference data) will require
rethinking and retooling of other pipeline components. We therefore
recommend that model pipelines be revisited holistically. This study
represents one attempt at diagnosing an update involving the NSVB system
of allometric models with a well-established Landsat-based modeling
approach, proposes potential solutions to uncovered challenges, and
identifies the overall relevance of these challenges to end-use
applications such as carbon monitoring and greenhouse gas inventories.

\hypertarget{acknowledgements}{%
\section{Acknowledgements}\label{acknowledgements}}

We would like to thank the USDA FIA program for their data sharing and
cooperation and the NYS Department of Environmental Conservation Office
of Climate Change for funding support.

\newpage{}

\hypertarget{references}{%
\section*{References}\label{references}}
\addcontentsline{toc}{section}{References}

\hypertarget{refs}{}
\begin{CSLReferences}{1}{0}
\leavevmode\vadjust pre{\hypertarget{ref-ayrey2021}{}}%
Ayrey, Elias, Daniel J. Hayes, John B. Kilbride, Shawn Fraver, John A.
Kershaw, Bruce D. Cook, and Aaron R. Weiskittel. 2021. {``Synthesizing
Disparate LiDAR and Satellite Datasets Through Deep Learning to Generate
Wall-to-Wall Regional Inventories for the Complex, Mixed-Species Forests
of the Eastern United States.''} \emph{Remote Sensing} 13 (24): 5113.
\url{https://doi.org/10.3390/rs13245113}.

\leavevmode\vadjust pre{\hypertarget{ref-exactextractr}{}}%
Baston, Daniel. 2022. \emph{{exactextractr}: Fast Extraction from Raster
Datasets Using Polygons}.
\url{https://CRAN.R-project.org/package=exactextractr}.

\leavevmode\vadjust pre{\hypertarget{ref-Bechtold2005}{}}%
Bechtold, William A, and Paul L Patterson. 2005. \emph{The Enhanced
Forest Inventory and Analysis Program--National Sampling Design and
Estimation Procedures}. Vol. 80. USDA Forest Service, Southern Research
Station. \url{https://doi.org/10.2737/SRS-GTR-80}.

\leavevmode\vadjust pre{\hypertarget{ref-beven1979}{}}%
Beven, K J, and M J Kirkby. 1979. {``A Physically Based, Variable
Contributing Area Model of Basin Hydrology / Un Modèle à Base Physique
de Zone d'appel Variable de l'hydrologie Du Bassin Versant.''}
\emph{Hydrological Sciences Bulletin} 24 (1): 43--69.
\url{https://doi.org/10.1080/02626667909491834}.

\leavevmode\vadjust pre{\hypertarget{ref-breiman2001}{}}%
Breiman, Leo. 2001. {``Random Forests.''} \emph{Machine Learning} 45
(1): 5--32. \url{https://doi.org/10.1023/a:1010933404324}.

\leavevmode\vadjust pre{\hypertarget{ref-brown2020}{}}%
Brown, Jesslyn F., Heather J. Tollerud, Christopher P. Barber, Qiang
Zhou, John L. Dwyer, James E. Vogelmann, Thomas R. Loveland, et al.
2020. {``Lessons Learned Implementing an Operational Continuous United
States National Land Change Monitoring Capability: The Land Change
Monitoring, Assessment, and Projection (LCMAP) Approach.''} \emph{Remote
Sensing of Environment} 238 (March): 111356.
\url{https://doi.org/10.1016/j.rse.2019.111356}.

\leavevmode\vadjust pre{\hypertarget{ref-IPCC2019}{}}%
Buendia, E, K Tanabe, A Kranjc, J Baasansuren, M Fukuda, S Ngarize, A
Osako, Y Pyrozhenko, P Shermanau, and S Federici. 2019. {``Refinement to
the 2006 IPCC Guidelines for National Greenhouse Gas Inventories.''}
\emph{IPCC: Geneva, Switzerland} 5: 194.

\leavevmode\vadjust pre{\hypertarget{ref-CEC}{}}%
CEC. 1997. \emph{Ecological Regions of North America: Toward a Common
Perspective}. Commission for Environmental Cooperation (Montr{é}al,
Qu{é}bec).; Secretariat.

\leavevmode\vadjust pre{\hypertarget{ref-Charnley2010}{}}%
Charnley, Susan, David Diaz, and Hannah Gosnell. 2010. {``Mitigating
Climate Change Through Small-Scale Forestry in the {USA}: Opportunities
and Challenges.''} \emph{Small-Scale Forestry} 9 (4): 445--62.
\url{https://doi.org/10.1007/s11842-010-9135-x}.

\leavevmode\vadjust pre{\hypertarget{ref-Chen1996}{}}%
Chen, Jing M. 1996. {``Evaluation of Vegetation Indices and a Modified
Simple Ratio for Boreal Applications.''} \emph{Canadian Journal of
Remote Sensing} 22 (3): 229--42.
\url{https://doi.org/10.1080/07038992.1996.10855178}.

\leavevmode\vadjust pre{\hypertarget{ref-Cocke2005}{}}%
Cocke, Allison E., Peter Z. Fulé, and Joseph E. Crouse. 2005.
{``Comparison of Burn Severity Assessments Using Differenced Normalized
Burn Ratio and Ground Data.''} \emph{International Journal of Wildland
Fire} 14: 189--98. \url{https://doi.org/10.1071/WF04010}.

\leavevmode\vadjust pre{\hypertarget{ref-cortes1995}{}}%
Cortes, Corinna, and Vladimir Vapnik. 1995. {``Support-Vector
Networks.''} \emph{Machine Learning} 20 (3): 273--97.
\url{https://doi.org/10.1007/bf00994018}.

\leavevmode\vadjust pre{\hypertarget{ref-Deo2016}{}}%
Deo, Ram K., Matthew B. Russell, Grant M. Domke, Christopher W. Woodall,
Michael J. Falkowski, and Warren B. Cohen. 2016. {``Using Landsat
Time-Series and LiDAR to Inform Aboveground Forest Biomass Baselines in
Northern Minnesota, USA.''} \emph{Canadian Journal of Remote Sensing} 43
(1): 28--47. \url{https://doi.org/10.1080/07038992.2017.1259556}.

\leavevmode\vadjust pre{\hypertarget{ref-dormann2018}{}}%
Dormann, Carsten F., Justin M. Calabrese, Gurutzeta Guillera-Arroita,
Eleni Matechou, Volker Bahn, Kamil Bartoń, Colin M. Beale, et al. 2018.
{``Model Averaging in Ecology: A Review of Bayesian,
Information{-}Theoretic, and Tactical Approaches for Predictive
Inference.''} \emph{Ecological Monographs} 88 (4): 485--504.
\url{https://doi.org/10.1002/ecm.1309}.

\leavevmode\vadjust pre{\hypertarget{ref-draper1997}{}}%
Draper, Norman R., and Yonghong (Fred) Yang. 1997. {``Generalization of
the Geometric Mean Functional Relationship.''} \emph{Computational
Statistics \& Data Analysis} 23 (3): 355--72.
\url{https://doi.org/10.1016/s0167-9473(96)00037-0}.

\leavevmode\vadjust pre{\hypertarget{ref-Dubayah2020}{}}%
Dubayah, Ralph, James Bryan Blair, Scott Goetz, Lola Fatoyinbo, Matthew
Hansen, Sean Healey, Michelle Hofton, et al. 2020. {``The Global
Ecosystem Dynamics Investigation: High-Resolution Laser Ranging of the
Earth's Forests and Topography.''} \emph{Science of Remote Sensing} 1
(June): 100002. \url{https://doi.org/10.1016/j.srs.2020.100002}.

\leavevmode\vadjust pre{\hypertarget{ref-Duncanson2010}{}}%
Duncanson, LI, KO Niemann, and MA Wulder. 2010. {``Integration of GLAS
and Landsat TM Data for Aboveground Biomass Estimation.''}
\emph{Canadian Journal of Remote Sensing} 36 (2): 129--41.
\url{https://doi.org/10.5589/m10-037}.

\leavevmode\vadjust pre{\hypertarget{ref-Dyer2006}{}}%
Dyer, James M. 2006. {``Revisiting the Deciduous Forests of Eastern
North America.''} \emph{BioScience} 56 (4): 341--52.
\url{https://doi.org/10.1641/0006-3568(2006)56\%5B341:RTDFOE\%5D2.0.CO;2}.

\leavevmode\vadjust pre{\hypertarget{ref-IPCC2006}{}}%
Eggleston, H S, L Buendia, K Miwa, T Ngara, and K Tanabe. 2006. {``2006
IPCC Guidelines for National Greenhouse Gas Inventories.''}

\leavevmode\vadjust pre{\hypertarget{ref-friedman2002}{}}%
Friedman, Jerome H. 2002. {``Stochastic Gradient Boosting.''}
\emph{Computational Statistics \& Data Analysis} 38 (4): 367--78.
\url{https://doi.org/10.1016/s0167-9473(01)00065-2}.

\leavevmode\vadjust pre{\hypertarget{ref-Gillenwater2007}{}}%
Gillenwater, Michael, Derik Broekhoff, Mark Trexler, Jasmine Hyman, and
Rob Fowler. 2007. {``Policing the Voluntary Carbon Market.''}
\emph{Nature Climate Change} 1 (711): 85--87.
\url{https://doi.org/10.1038/climate.2007.58}.

\leavevmode\vadjust pre{\hypertarget{ref-gray2012}{}}%
Gray, Andrew, Thomas Brandeis, John Shaw, William McWilliams, and
Patrick Miles. 2012. {``Forest Inventory and Analysis Database of the
United States of America (FIA).''} \emph{Biodiversity \& Ecology} 4
(September): 225--31. \url{https://doi.org/10.7809/b-e.00079}.

\leavevmode\vadjust pre{\hypertarget{ref-Hall2006}{}}%
Hall, R. J., R. S. Skakun, E. J. Arsenault, and B. S. Case. 2006.
{``Modeling Forest Stand Structure Attributes Using Landsat ETM+ Data:
Application to Mapping of Aboveground Biomass and Stand Volume.''}
\emph{Forest Ecology and Management} 225 (1--3): 378--90.
\url{https://doi.org/10.1016/j.foreco.2006.01.014}.

\leavevmode\vadjust pre{\hypertarget{ref-heath2009}{}}%
Heath, L S, Mark H Hansen, James E Smith, W Brad Smith, and Patrick D
Miles. 2009. {``Investigation into Calculating Tree Biomass and Carbon
in the FIADB Using a Biomass Expansion Factor Approach.''} In \emph{2008
Forest Inventory and Analysis (FIA) Symposium}, edited by Will
McWilliams, Gretchen Moisen, and Ray Czaplewski, 21--23. Fort Collins,
CO: U.S. Department of Agriculture, Forest Service, Rocky Mountain
Research Station; U.S. Department of Agriculture, Forest Service, Rocky
Mountain Research Station.

\leavevmode\vadjust pre{\hypertarget{ref-terra}{}}%
Hijmans, Robert J. 2023. \emph{{terra}: Spatial Data Analysis}.
\url{https://rspatial.org/}.

\leavevmode\vadjust pre{\hypertarget{ref-huang2019}{}}%
Huang, Wenli, Katelyn Dolan, Anu Swatantran, Kristofer Johnson, Hao
Tang, Jarlath O'Neil-Dunne, Ralph Dubayah, and George Hurtt. 2019.
{``High-Resolution Mapping of Aboveground Biomass for Forest Carbon
Monitoring System in the Tri-State Region of Maryland, Pennsylvania and
Delaware, USA.''} \emph{Environmental Research Letters} 14 (9): 095002.
\url{https://doi.org/10.1088/1748-9326/ab2917}.

\leavevmode\vadjust pre{\hypertarget{ref-hudak2020}{}}%
Hudak, Andrew T, Patrick A Fekety, Van R Kane, Robert E Kennedy, Steven
K Filippelli, Michael J Falkowski, Wade T Tinkham, et al. 2020. {``A
Carbon Monitoring System for Mapping Regional, Annual Aboveground
Biomass Across the Northwestern USA.''} \emph{Environmental Research
Letters} 15 (9): 095003. \url{https://doi.org/10.1088/1748-9326/ab93f9}.

\leavevmode\vadjust pre{\hypertarget{ref-ji2006}{}}%
Ji, Lei, and Kevin Gallo. 2006. {``An Agreement Coefficient for Image
Comparison.''} \emph{Photogrammetric Engineering \& Remote Sensing} 72
(7): 823--33. \url{https://doi.org/10.14358/pers.72.7.823}.

\leavevmode\vadjust pre{\hypertarget{ref-johnson2022}{}}%
Johnson, Lucas K, Michael J Mahoney, Eddie Bevilacqua, Stephen V
Stehman, Grant M Domke, and Colin M Beier. 2022. {``Fine-Resolution
Landscape-Scale Biomass Mapping Using a Spatiotemporal Patchwork of
LiDAR Coverages.''} \emph{International Journal of Applied Earth
Observation and Geoinformation} 114 (November): 103059.
\url{https://doi.org/10.1016/j.jag.2022.103059}.

\leavevmode\vadjust pre{\hypertarget{ref-johnson2023}{}}%
Johnson, Lucas K, Michael J Mahoney, Madeleine L Desrochers, and Colin M
Beier. 2023. {``Mapping Historical Forest Biomass for Stock-Change
Assessments at Parcel to Landscape Scales.''} \emph{Forest Ecology and
Management} 546 (October): 121348.
\url{https://doi.org/10.1016/j.foreco.2023.121348}.

\leavevmode\vadjust pre{\hypertarget{ref-Jordan1969}{}}%
Jordan, Carl F. 1969. {``Derivation of Leaf-Area Index from Quality of
Light on the Forest Floor.''} \emph{Ecology} 50 (4): 663--66.
\url{https://doi.org/10.2307/1936256}.

\leavevmode\vadjust pre{\hypertarget{ref-Kaarakka2021}{}}%
Kaarakka, Lilli, Meredith Cornett, Grant Domke, Todd Ontl, and Laura E.
Dee. 2021. {``Improved Forest Management as a Natural Climate Solution:
A Review.''} \emph{Ecological Solutions and Evidence} 2 (3).
\url{https://doi.org/10.1002/2688-8319.12090}.

\leavevmode\vadjust pre{\hypertarget{ref-kernlab}{}}%
Karatzoglou, Alexandros, Alex Smola, Kurt Hornik, and Achim Zeileis.
2004. {``{kernlab} -- an {S4} Package for Kernel Methods in {R}.''}
\emph{Journal of Statistical Software} 11 (9): 1--20.
\url{https://doi.org/10.18637/jss.v011.i09}.

\leavevmode\vadjust pre{\hypertarget{ref-Kauth1976}{}}%
Kauth, Richard J., and G. S. P. Thomas. 1976. {``The Tasselled Cap - a
Graphic Description of the Spectral-Temporal Development of Agricultural
Crops as Seen by Landsat.''} In \emph{Symposium on Machine Processing of
Remotely Sensed Data}, 159.

\leavevmode\vadjust pre{\hypertarget{ref-Guolin2017}{}}%
Ke, Guolin, Qi Meng, Thomas Finley, Taifeng Wang, Wei Chen, Weidong Ma,
Qiwei Ye, and Tie-Yan Liu. 2017. {``{LightGBM: A Highly Efficient
Gradient Boosting Decision Tree}.''} In \emph{Advances in Neural
Information Processing Systems}, edited by I. Guyon, U. V. Luxburg, S.
Bengio, H. Wallach, R. Fergus, S. Vishwanathan, and R. Garnett. Vol. 30.
Curran Associates, Inc.
\url{https://proceedings.neurips.cc/paper/2017/file/6449f44a102fde848669bdd9eb6b76fa-Paper.pdf}.

\leavevmode\vadjust pre{\hypertarget{ref-kennedy2018}{}}%
Kennedy, Robert E, Janet Ohmann, Matt Gregory, Heather Roberts, Zhiqiang
Yang, David M Bell, Van Kane, et al. 2018. {``An Empirical, Integrated
Forest Biomass Monitoring System.''} \emph{Environmental Research
Letters} 13 (2): 025004. \url{https://doi.org/10.1088/1748-9326/aa9d9e}.

\leavevmode\vadjust pre{\hypertarget{ref-kennedy2010}{}}%
Kennedy, Robert E, Zhiqiang Yang, and Warren B Cohen. 2010. {``Detecting
Trends in Forest Disturbance and Recovery Using Yearly Landsat Time
Series: 1. LandTrendr {\textemdash} Temporal Segmentation Algorithms.''}
\emph{Remote Sensing of Environment} 114 (12): 2897--2910.
\url{https://doi.org/10.1016/j.rse.2010.07.008}.

\leavevmode\vadjust pre{\hypertarget{ref-kennedy2018a}{}}%
Kennedy, Robert E, Zhiqiang Yang, Noel Gorelick, Justin Braaten, Lucas
Cavalcante, Warren Cohen, and Sean Healey. 2018. {``Implementation of
the LandTrendr Algorithm on Google Earth Engine.''} \emph{Remote
Sensing} 10 (5): 691. \url{https://doi.org/10.3390/rs10050691}.

\leavevmode\vadjust pre{\hypertarget{ref-Kriegler1969}{}}%
Kriegler, F. J., W. A. Malila, R. F. Nalepka, and W. Richardson. 1969.
{``{Preprocessing Transformations and Their Effects on Multispectral
Recognition}.''} In \emph{Remote Sensing of Environment, VI}, 97.

\leavevmode\vadjust pre{\hypertarget{ref-Lorimer2001}{}}%
Lorimer, Craig G. 2001. {``Historical and Ecological Roles of
Disturbance in Eastern North American Forests: 9,000 Years of Change.''}
\emph{Wildlife Society Bulletin (1973-2006)} 29 (2): 425--39.
\url{http://www.jstor.org/stable/3784167}.

\leavevmode\vadjust pre{\hypertarget{ref-Lu2005}{}}%
Lu, Dengsheng. 2005. {``Aboveground Biomass Estimation Using Landsat TM
Data in the Brazilian Amazon.''} \emph{International Journal of Remote
Sensing} 26 (12): 2509--25.
\url{https://doi.org/10.1080/01431160500142145}.

\leavevmode\vadjust pre{\hypertarget{ref-Lu2006}{}}%
---------. 2006. {``The Potential and Challenge of Remote Sensing‐based
Biomass Estimation.''} \emph{International Journal of Remote Sensing} 27
(7): 1297--1328. \url{https://doi.org/10.1080/01431160500486732}.

\leavevmode\vadjust pre{\hypertarget{ref-mahoney2023}{}}%
Mahoney, Michael J. 2023. {``Waywiser: Ergonomic Methods for Assessing
Spatial Models.''} \url{https://doi.org/10.48550/arXiv.2303.11312}.

\leavevmode\vadjust pre{\hypertarget{ref-terrainr}{}}%
Mahoney, Michael J, Colin M Beier, and Aidan C Ackerman. 2022.
{``{terrainr}: An r Package for Creating Immersive Virtual
Environments.''} \emph{Journal of Open Source Software} 7 (69): 4060.
\url{https://doi.org/10.21105/joss.04060}.

\leavevmode\vadjust pre{\hypertarget{ref-mahoney2022}{}}%
Mahoney, Michael J, Lucas K Johnson, Abigail Z Guinan, and Colin M
Beier. 2022. {``Classification and Mapping of Low-Statured Shrubland
Cover Types in Post-Agricultural Landscapes of the US Northeast.''}
\emph{International Journal of Remote Sensing} 43 (19-24): 7117--38.
\url{https://doi.org/10.1080/01431161.2022.2155086}.

\leavevmode\vadjust pre{\hypertarget{ref-massey1951}{}}%
Massey Jr, Frank J. 1951. {``The Kolmogorov-Smirnov Test for Goodness of
Fit.''} \emph{Journal of the American Statistical Association}, 68--78.
\url{https://doi.org/10.2307/2280095}.

\leavevmode\vadjust pre{\hypertarget{ref-mcroberts2011}{}}%
McRoberts, Ronald E. 2011. {``Satellite Image-Based Maps: Scientific
Inference or Pretty Pictures?''} \emph{Remote Sensing of Environment}
115 (2): 715--24. \url{https://doi.org/10.1016/j.rse.2010.10.013}.

\leavevmode\vadjust pre{\hypertarget{ref-montero2022}{}}%
Montero, D., C. Aybar, M. D. Mahecha, and S. Wieneke. 2022. {``SPECTRAL:
AWESOME SPECTRAL INDICES DEPLOYED VIA THE GOOGLE EARTH ENGINE JAVASCRIPT
API.''} \emph{The International Archives of the Photogrammetry, Remote
Sensing and Spatial Information Sciences} XLVIII-4/W1-2022 (August):
301--6.
\url{https://doi.org/10.5194/isprs-archives-xlviii-4-w1-2022-301-2022}.

\leavevmode\vadjust pre{\hypertarget{ref-NOAA}{}}%
NOAA National Centers for Environmental Information. 2022. {``Climate at
a Glance: Statewide Mapping.''} \url{https://www.ncdc.noaa.gov/cag/}.

\leavevmode\vadjust pre{\hypertarget{ref-omernik2014}{}}%
Omernik, James M., and Glenn E. Griffith. 2014. {``Ecoregions of the
Conterminous United States: Evolution of a Hierarchical Spatial
Framework.''} \emph{Environmental Management} 54 (6): 1249--66.
\url{https://doi.org/10.1007/s00267-014-0364-1}.

\leavevmode\vadjust pre{\hypertarget{ref-Pflugmacher2012}{}}%
Pflugmacher, Dirk, Warren B. Cohen, and Robert E. Kennedy. 2012.
{``Using Landsat-Derived Disturbance History (1972--2010) to Predict
Current Forest Structure.''} \emph{Remote Sensing of Environment} 122
(July): 146--65. \url{https://doi.org/10.1016/j.rse.2011.09.025}.

\leavevmode\vadjust pre{\hypertarget{ref-Powell2010}{}}%
Powell, Scott L., Warren B. Cohen, Sean P. Healey, Robert E. Kennedy,
Gretchen G. Moisen, Kenneth B. Pierce, and Janet L. Ohmann. 2010.
{``Quantification of Live Aboveground Forest Biomass Dynamics with
Landsat Time-Series and Field Inventory Data: A Comparison of Empirical
Modeling Approaches.''} \emph{Remote Sensing of Environment} 114 (5):
1053--68. \url{https://doi.org/10.1016/j.rse.2009.12.018}.

\leavevmode\vadjust pre{\hypertarget{ref-PRISM}{}}%
PRISM Climate Group. 2022. {``PRISM Climate Data.''}
\url{https://prism.oregonstate.edu}.

\leavevmode\vadjust pre{\hypertarget{ref-fia_pop}{}}%
Pugh, Scott A., Jeffery A. Turner, Elizabeth A. Burrill, and Winnie
David. 2018. {``The Forest Inventory and Analysis Database: Population
Estimation User Guide.''}
\url{https://www.fia.fs.usda.gov/library/database-documentation/current/ver80/FIADB\%20Population\%20Estimation\%20user\%20guide_11_2018_final_revised_02_2019.pdf}.

\leavevmode\vadjust pre{\hypertarget{ref-Quegan2019}{}}%
Quegan, Shaun, Thuy Le Toan, Jerome Chave, Jorgen Dall, Jean-François
Exbrayat, Dinh Ho Tong Minh, Mark Lomas, et al. 2019. {``The European
Space Agency BIOMASS Mission: Measuring Forest Above-Ground Biomass from
Space.''} \emph{Remote Sensing of Environment} 227 (June): 44--60.
\url{https://doi.org/10.1016/j.rse.2019.03.032}.

\leavevmode\vadjust pre{\hypertarget{ref-R}{}}%
R Core Team. 2023. \emph{R: A Language and Environment for Statistical
Computing}. Vienna, Austria: R Foundation for Statistical Computing.
\url{https://www.R-project.org/}.

\leavevmode\vadjust pre{\hypertarget{ref-Riemann2014}{}}%
Riemann, Rachel I., Barry T. Wilson, Andrew J. Lister, Oren Cook, and
Sierra. Crane-Murdoch. 2014. \emph{Modeled Distributions of 12 Tree
Species in New York}. U.S. Department of Agriculture, Forest Service,
Northern Research Station. \url{https://doi.org/10.2737/nrs-rmap-5}.

\leavevmode\vadjust pre{\hypertarget{ref-riemann2010}{}}%
Riemann, Rachel I., Barry Tyler Wilson, Andrew Lister, and Sarah Parks.
2010. {``An Effective Assessment Protocol for Continuous Geospatial
Datasets of Forest Characteristics Using USFS Forest Inventory and
Analysis (FIA) Data.''} \emph{Remote Sensing of Environment} 114 (10):
2337--52. \url{https://doi.org/10.1016/j.rse.2010.05.010}.

\leavevmode\vadjust pre{\hypertarget{ref-roy2016}{}}%
Roy, D. P., V. Kovalskyy, H. K. Zhang, E. F. Vermote, L. Yan, S. S.
Kumar, and A. Egorov. 2016. {``Characterization of Landsat-7 to
Landsat-8 Reflective Wavelength and Normalized Difference Vegetation
Index Continuity.''} \emph{Remote Sensing of Environment} 185
(November): 57--70. \url{https://doi.org/10.1016/j.rse.2015.12.024}.

\leavevmode\vadjust pre{\hypertarget{ref-lightgbm}{}}%
Shi, Yu, Guolin Ke, Damien Soukhavong, James Lamb, Qi Meng, Thomas
Finley, Taifeng Wang, et al. 2022. \emph{Lightgbm: Light Gradient
Boosting Machine}. \url{https://github.com/Microsoft/LightGBM}.

\leavevmode\vadjust pre{\hypertarget{ref-simard2011}{}}%
Simard, Marc, Naiara Pinto, Joshua B. Fisher, and Alessandro Baccini.
2011. {``Mapping Forest Canopy Height Globally with Spaceborne Lidar.''}
\emph{Journal of Geophysical Research} 116 (G4).
\url{https://doi.org/10.1029/2011jg001708}.

\leavevmode\vadjust pre{\hypertarget{ref-stehman2019}{}}%
Stehman, Stephen V., and Giles M. Foody. 2019. {``Key Issues in Rigorous
Accuracy Assessment of Land Cover Products.''} \emph{Remote Sensing of
Environment} 231 (September): 111199.
\url{https://doi.org/10.1016/j.rse.2019.05.018}.

\leavevmode\vadjust pre{\hypertarget{ref-USGS_dem}{}}%
U.S. Geological Survey. 2019. {``{3D} Elevation Program 1-Meter
Resolution Digital Elevation Model.''}
\url{https://www.usgs.gov/the-national-map-data-delivery}.

\leavevmode\vadjust pre{\hypertarget{ref-Census}{}}%
US Census Bureau. 2013. {``TIGER/Line Shapefiles.''}
\url{https://www.census.gov/geographies/mapping-files/time-series/geo/tiger-line-file.html}.

\leavevmode\vadjust pre{\hypertarget{ref-landsat2018}{}}%
USGS. 2018. {``Landsat Collections.''} {US} Geological Survey.
\url{https://doi.org/10.3133/fs20183049}.

\leavevmode\vadjust pre{\hypertarget{ref-nsvb_ind}{}}%
Virginia Tech. 2023a. {``Overview of the National Scale Volume and
Biomass Estimators (NSVB): State Report for Indiana.''}
\url{https://charcoal2.cnre.vt.edu/nsvb_factsheets/Reports/State/Indiana.html}.

\leavevmode\vadjust pre{\hypertarget{ref-nsvb_nys}{}}%
---------. 2023b. {``Overview of the National Scale Volume and Biomass
Estimators (NSVB): State Report for New York.''}
\url{https://charcoal2.cnre.vt.edu/nsvb_factsheets/Reports/State/New\%20York.html}.

\leavevmode\vadjust pre{\hypertarget{ref-nsvb_wash}{}}%
---------. 2023c. {``Overview of the National Scale Volume and Biomass
Estimators (NSVB): State Report for Washington.''}
\url{https://charcoal2.cnre.vt.edu/nsvb_factsheets/Reports/State/Washington.html}.

\leavevmode\vadjust pre{\hypertarget{ref-tigris}{}}%
Walker, Kyle. 2023. \emph{{tigris}: Load Census TIGER/Line Shapefiles}.
\url{https://CRAN.R-project.org/package=tigris}.

\leavevmode\vadjust pre{\hypertarget{ref-Westfall2023}{}}%
Westfall, James A., John W. Coulston, Andrew N. Gray, John D. Shaw,
Philip J. Radtke, David M. Walker, Aaron R. Weiskittel, et al. 2023.
\emph{A National-Scale Tree Volume, Biomass, and Carbon Modeling System
for the United States}. U.S. Department of Agriculture, Forest Service.
\url{https://doi.org/10.2737/wo-gtr-104}.

\leavevmode\vadjust pre{\hypertarget{ref-Whitney1994}{}}%
Whitney, Gordon G. 1994. \emph{From Coastal Wilderness to Fruited Plain:
A History of Environmental Change in Temperate North America from 1500
to the Present}. Cambridge, United Kingdom: Cambridge University Press.

\leavevmode\vadjust pre{\hypertarget{ref-willmott2011}{}}%
Willmott, Cort J., Scott M. Robeson, and Kenji Matsuura. 2011. {``A
Refined Index of Model Performance.''} \emph{International Journal of
Climatology} 32 (13): 2088--94. \url{https://doi.org/10.1002/joc.2419}.

\leavevmode\vadjust pre{\hypertarget{ref-Wilson2012}{}}%
Wilson, B. Tyler, Andrew J. Lister, and Rachel I. Riemann. 2012. {``A
Nearest-Neighbor Imputation Approach to Mapping Tree Species over Large
Areas Using Forest Inventory Plots and Moderate Resolution Raster
Data.''} \emph{Forest Ecology and Management} 271 (May): 182--98.
\url{https://doi.org/10.1016/j.foreco.2012.02.002}.

\leavevmode\vadjust pre{\hypertarget{ref-Wilson2013}{}}%
Wilson, B. Tyler, Andrew J. Lister, Rachel I. Riemann, and Douglas M.
Griffith. 2013. {``Live Tree Species Basal Area of the Contiguous United
States (2000-2009).''} \emph{Forest Service Research Data Archive}.
Forest Service Research Data Archive.
\url{https://doi.org/10.2737/rds-2013-0013}.

\leavevmode\vadjust pre{\hypertarget{ref-wolpert1992}{}}%
Wolpert, David H. 1992. {``Stacked Generalization.''} \emph{Neural
Networks} 5 (2): 241--59.
\url{https://doi.org/10.1016/s0893-6080(05)80023-1}.

\leavevmode\vadjust pre{\hypertarget{ref-woodall2011}{}}%
Woodall, Christopher W., G. M. Domke, D. W. Macfarlane, and C. M.
Oswalt. 2011. {``Comparing Field- and Model-Based Standing Dead Tree
Carbon Stock Estimates Across Forests of the US.''} \emph{Forestry} 85
(1): 125--33. \url{https://doi.org/10.1093/forestry/cpr065}.

\leavevmode\vadjust pre{\hypertarget{ref-woodall2011a}{}}%
Woodall, Christopher W., Linda S. Heath, Grant M. Domke, and Michael C.
Nichols. 2011. {``Methods and Equations for Estimating Aboveground
Volume, Biomass, and Carbon for Trees in the u.s. Forest Inventory,
2010.''} U.S. Department of Agriculture, Forest Service, Northern
Research Station. \url{https://doi.org/10.2737/nrs-gtr-88}.

\leavevmode\vadjust pre{\hypertarget{ref-wright2017}{}}%
Wright, Marvin N., and Andreas Ziegler. 2017. {``{\textbf{Ranger}}: A
Fast Implementation of Random Forests for High Dimensional Data in
{\emph{C++}} and {\emph{R}}.''} \emph{Journal of Statistical Software}
77 (1). \url{https://doi.org/10.18637/jss.v077.i01}.

\leavevmode\vadjust pre{\hypertarget{ref-zheng2007}{}}%
Zheng, D, L S Heath, and M J Ducey. 2007. {``Forest Biomass Estimated
from MODIS and FIA Data in the Lake States: MN, WI and MI, USA.''}
\emph{Forestry} 80 (3): 265--78.
\url{https://doi.org/10.1093/forestry/cpm015}.

\leavevmode\vadjust pre{\hypertarget{ref-zhu2014}{}}%
Zhu, Zhe, and Curtis E Woodcock. 2014. {``Continuous Change Detection
and Classification of Land Cover Using All Available Landsat Data.''}
\emph{Remote Sensing of Environment} 144 (March): 152--71.
\url{https://doi.org/10.1016/j.rse.2014.01.011}.

\end{CSLReferences}

\end{document}